# A review of dynamics design methods for high-speed and high-precision CNC machine tool feed systems

Xuesong Wang[1], Dongsheng Zhang[1]*

（[1]Xi'an Jiaotong University, Xi'an, Shaanxi, CN）

(*Correspondence: Dongsheng Zhang)

**Abstract:**

With the continuous development of the manufacturing industry, there is an increasing demand for high-speed and high-precision performance in CNC machine tool feed systems. The quality of the design method is a critical factor in determining its dynamic performance. Understanding the evolution mechanism of the feed system's dynamic performance is crucial for enhancing its dynamic characteristics and improving design methods. However, few studies have comprehensively reviewed previous related research. To fill this gap, this paper reviews the research achievements and progress related to the optimization design methods of CNC machine tool feed systems and the mechanisms influencing their dynamic performance. It comprehensively and deeply elucidates the current feed system dynamic design research status from three aspects: subsystem optimization, subsystem coupling mechanism, and dynamic matching design. In conclusion, this paper exploratory proposes the development direction for the integrated dynamic optimization design of CNC machine tool feed systems, providing potential insights for researchers in related fields.

## 1 Introduction

CNC machine tools are a strategic resource in developing high-end manufacturing industries. This is especially prominent as manufacturing industries move toward high dimensional accuracy and surface quality, increasing the feed system requirements regarding speed and precision performance [1–6].In engineering practice, the performance metrics of feed systems are typically time-domain indicators such as positioning accuracy and repeat positioning accuracy [7–10]. Engineers usually rely on empirical analogy designs to meet static accuracy requirements and employ methods such as lightweight, structural optimization, and assembly process optimization [11–24] to enhance the natural frequency of the feed system and perform inertia matching to improve dynamic accuracy [25–31]. However, these design approaches often fail to directly indicate whether the design can meet the dimensional accuracy and surface quality requirements of the machined workpiece [32,33].

This paper argues that the evolution mechanism of CNC machine tool dynamic performance should be analyzed from a dynamics perspective to develop optimization design schemes that directly address accuracy requirements. The dynamic characteristics of the CNC machine tool feed system can be analyzed using the schematic diagram shown in Figure 1. The system comprises four sub-systems: motion process $R(s)$, control system $G_c(s)$, motor $G_a(s)$, and mechanical body $G_p(s)$. The position error $\Delta e(t)$ is shown in Equation (1). $\mathcal{L}^{-1}$ is the inverse Laplace transform. Analysis reveals that the position error of the feed system is influenced by the comprehensive effects of the four major subsystems of the feed system [34–36]. The coupling interactions among the subsystems within the CNC machine tool feed system are highly complex [37–39], and its performance is affected by the integrated influence of these subsystems [40–44]. Current optimization design methods for CNC machine tool feed systems have not fully accounted for the complex coupling interactions between subsystems, making it difficult to maximize the dynamic performance of the feed system [45–51].

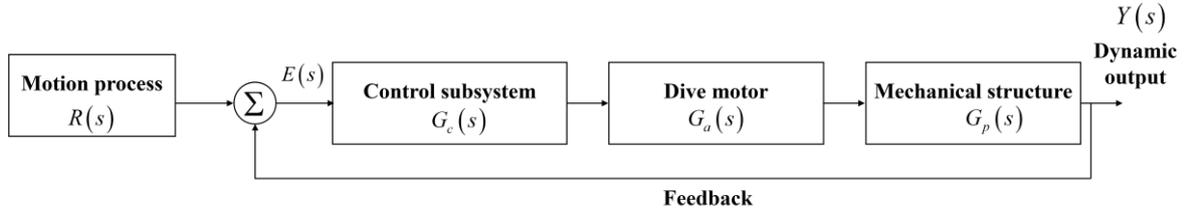

Figure 1 Block diagram of servo feed system components

$$\Delta e(t) = \mathcal{L}^{-1}\left[E(s)\right] = \mathcal{L}^{-1}\left[\frac{R(s)}{1+G_c(s)G_a(s)G_p(s)}\right] \quad (1)$$

To address these issues, this paper reviews and analyzes the existing design research on CNC machine tool feed systems. The structure of this paper is illustrated in Figure 2. First, most of the current performance optimization research for feed systems focuses on optimizing the performance of subsystems. However, the optimal performance of subsystems does not necessarily lead to the optimal overall performance of the feed system. Therefore, in the dynamic optimization design of feed systems, the integrated impact of subsystems should be considered. Second, this paper organizes and analyzes the research achievements on the coupling mechanisms of subsystems within feed systems. Following this, it analyzes the existing matching design methods for subsystems. Finally, the paper explores potential future development directions for dynamic design methods of feed systems.

This paper is organized as follows: Chapter 1 reviews the design optimization of the sub-systems, while studies on the coupling mechanism between sub-systems of the CNC machine tool feed system were covered in Chapter 2. In Chapter 3, the available studies on dynamical design matching were explained in detail. Finally, Chapter 4 provides a summary and exploratory proposes an outlook on the future dynamical design methods of feed systems of high-speed and high-precision CNC machine tools.

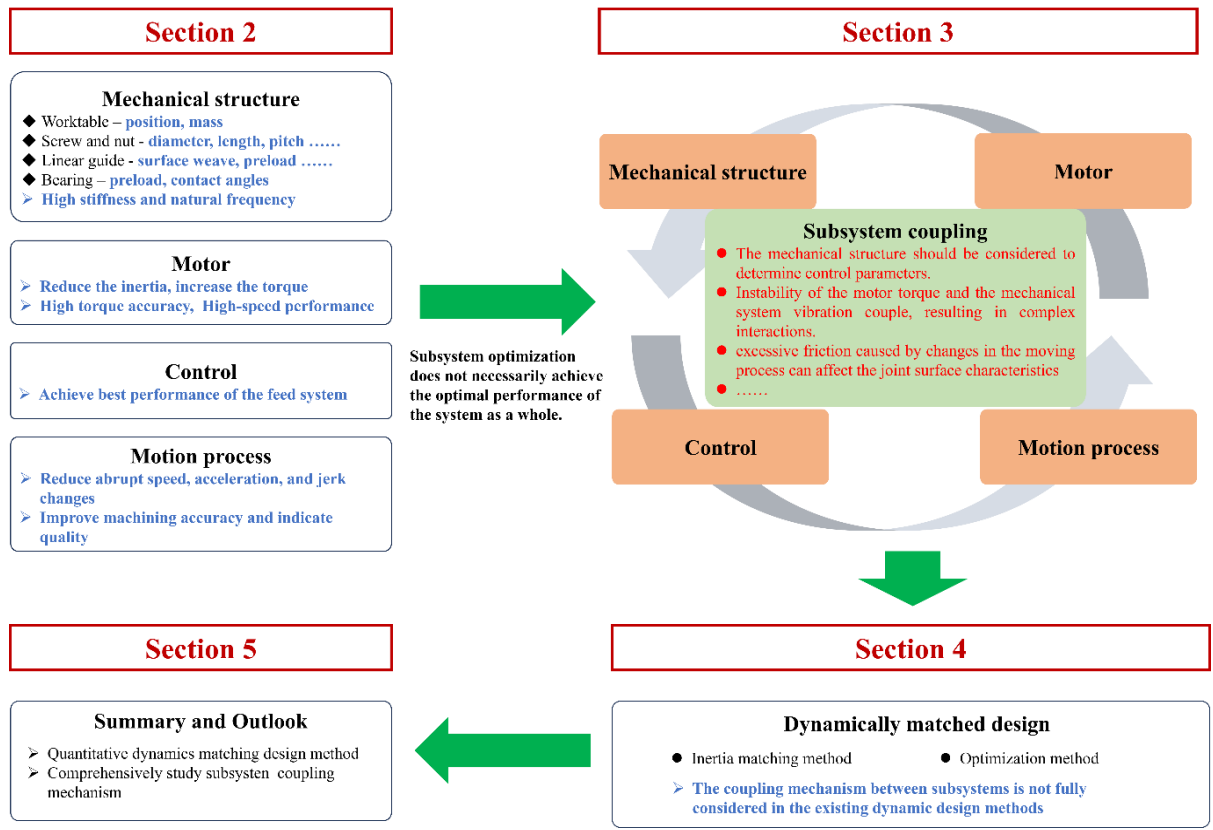

Figure 2 Content diagram of this paper

## 2 Performance optimization study of sub-systems

As previously mentioned, the feed system generally consists of four major subsystems: the mechanical system, motor, control system, and motion process. Optimizing the design of each subsystem can effectively enhance the dynamic performance of the feed system. The mechanical system should have high stiffness and natural frequency to ensure good response speed and vibration resistance. The motor should feature characteristics like lightweight and high torque density. The control system should be continuously optimized to maximize the dynamic characteristics of the electromechanical system. In motion process design, speed planning and interpolation algorithms should be optimized to ensure machining accuracy.

## 2.1 Mechanical structure optimization

In the mechanical system of the CNC machine tool feed system, the structural parameters of components such as the worktable, guide rails, bearings, ball screws, and nuts are key factors affecting its dynamic performance [52] (as shown in Table 1). Existing research has optimized the structural parameters of these components to continuously enhance their stiffness and natural frequency [11–17,24], ensuring good response speed and vibration resistance of the mechanical structure [44,53–57].

Table 1 Mechanical parameters that affect the performance of the feed system

| Component | Impact factor | Impact overview |
|---|---|---|
| Worktable | Position | The farther away the table is from the fixed end, the smaller the stiffness. (Figure 4, Figure 5, Figure 6) |
|  | Lightweight | The lightweight design of the workbench is carried out to increase its natural frequency and further improve the high-speed and high-precision performance of the CNC machine tool. |
| Screw-nut | Nominal diameter, length, and pitch | With the decrease of nominal diameter and pitch and the increase of length, the variation range of natural frequency increases obviously. (Figure 7, Figure 8) |
|  | Load and preload | With the increase of preload, the stiffness of the feed system increases. （Figure 9, Figure 10） |
|  | Small ball dimension and the phase | The vibration could be reduced by decreasing the small ball dimension and the phase difference between the ball groups. (Figure 11) |
|  | Nominal contact angle of ball screw nut pair | The larger the nominal contact angle is, the larger the fundamental resonance frequency is and the smaller the primary harmonic amplitude is. (Figure 12, Figure 13) |
| Linear guide | Load and preload | The frequency response of rolling linear guides significantly changes under specific cutting forces and vertical loads. (Figure 14, Figure 15) |
|  | Surface weave | The suitable surface weave can reduce guide friction, alleviating the guide stick-slip phenomenon. (Figure 16, Figure 17) |
|  | Profile curve | The guide profile curve directly affected the straightness and guide motion angular error. (Figure 18~Figure 20) |
| Bearing | Rotational speed and external load | Liu et al. analyzed the effects of external load , and rotational speed. The results have shown that the elastic deformation of rollers and raceways significantly affects their vibration characteristics.(Figure 21~Figure 23) |
| | Nominal contact angle | The larger the nominal contact angle is, the larger the fundamental resonance frequency is and the smaller the primary harmonic amplitude is. (Figure 26) |
| | Radial clearance | The damping ratio decreased with the increase in clearance. (Figure 25) |

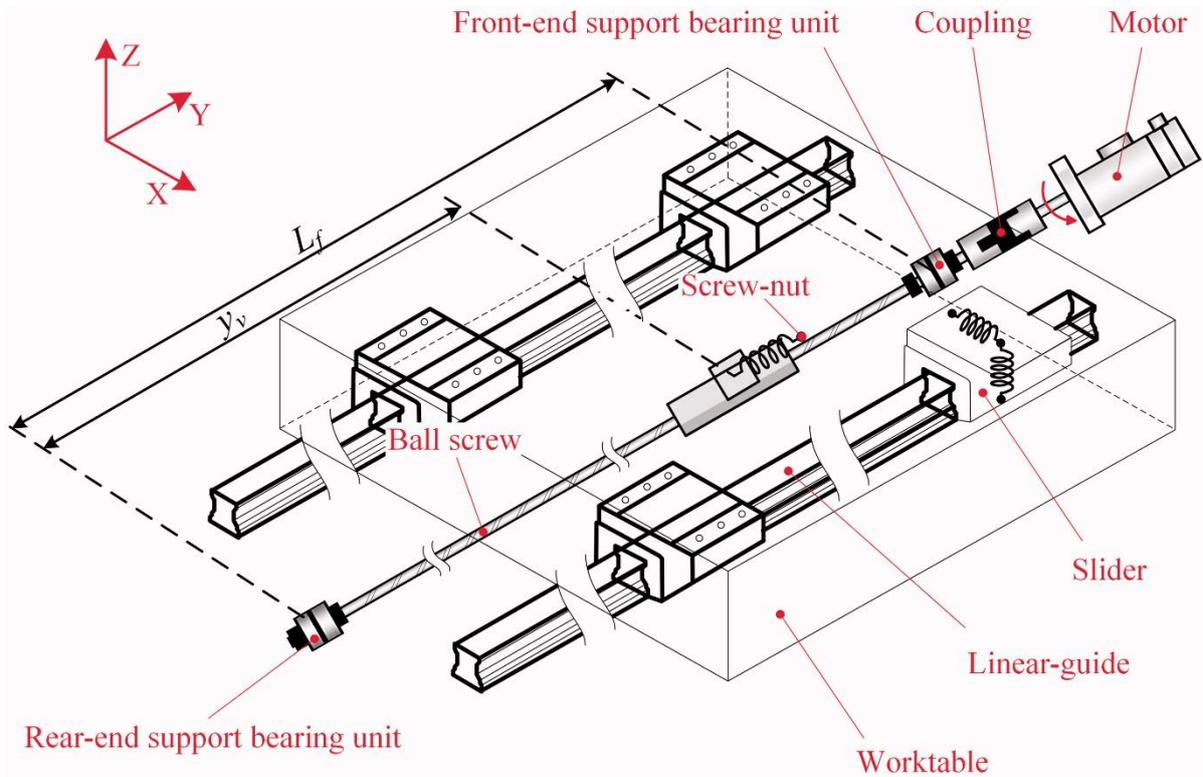

Figure 3 Schematic view of a ball-screw feed system [52]

### 2.1.1 Worktable

For machine tool feed systems utilizing ball screw drives, the table position changes significantly as the machining process is carried out, with the axial and torsional stiffness of the screw varying [17,58–60]. For a ball screw feed system (BSFS) with a fixed support installation method, the farther the sliding platform is from the fixed end, the lower the resonance frequency and the poorer the system stability. That is, short-stroke BSFS has higher positioning accuracy [30,61]. Moreover, other researchers have optimized the structure and materials to lighten the worktable, thereby increasing its natural frequency and further improving the high-speed and high-precision performance of the CNC machine tool feed system [27–29,62].

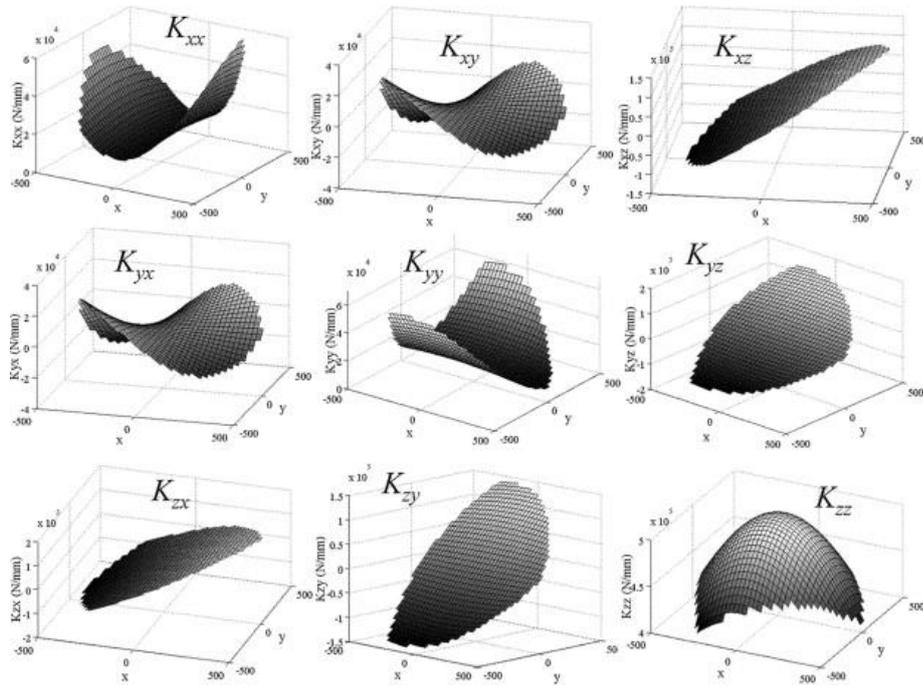

Figure 4 Static stiffness maps for a given altitude z [54]

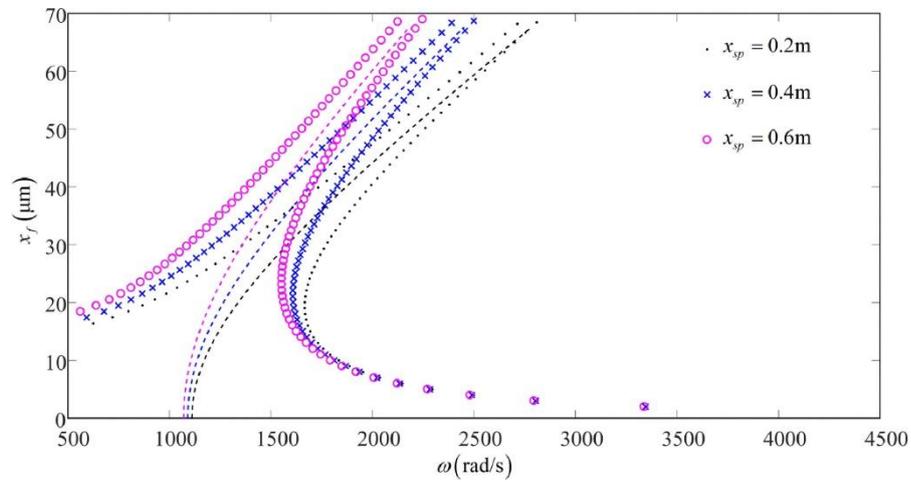

Figure 5 Fundamental response curves at different sliding platform positions [30]

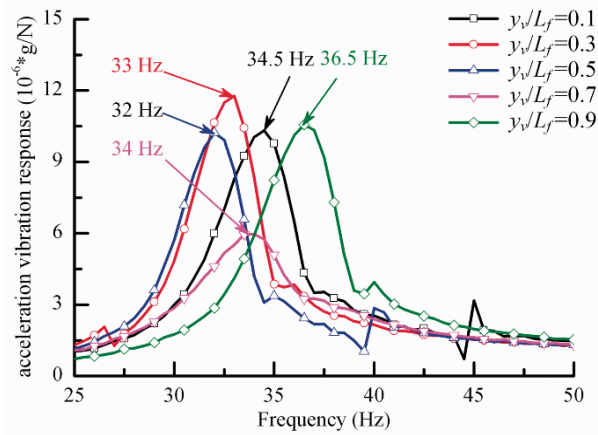

Figure 6 Experimental results of the frequency response under different worktable positions [52]

### 2.1.2 Screw and nut

The study by Zhang et al. [52] showed that the dynamics of ball screw feed drive systems change significantly throughout the working stroke. The nominal diameter, length, and screw pitches of the ball screw significantly affect the system's natural frequency. As the nominal diameter decreases and the length increases, the variation amplitude of the natural frequency increases noticeably. When designing the ball screw feed system, for more uniform dynamic characteristics, the ratio of the screw length to its nominal diameter should be limited to 50.

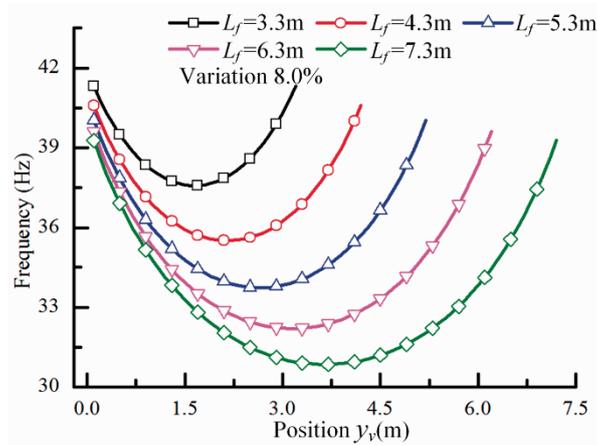

Figure 7 Variation of the natural frequency for different screw lengths [52]

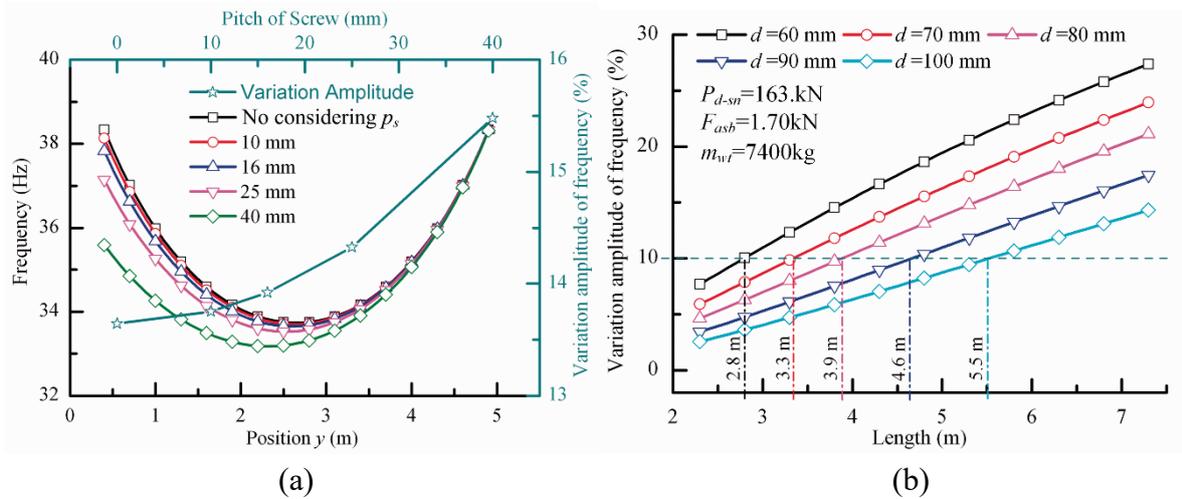

Figure 8 (a) Variation and amplitude of natural frequency for different screw pitches; (b) Variation amplitude of the natural frequency for different ratios of length to diameter of the screw-shaft [52]

Xu et al. [63] conducted a ball passage vibration analysis of the recirculating ball screw mechanism. It showed that, with the increase in external load and internal preload, the vibration amplitude and the peak-to-peak value of the dynamic response of the ball passage increased. Zhang et al. [22] developed a new dynamics model with a two-end fixed ball screw feed system. The effect of the screw preload force was accounted for, and both theoretical and experimental results showed that the screw preload force significantly affected the screw-nut contact state. As such, it caused a significant change in the natural frequency of the system.

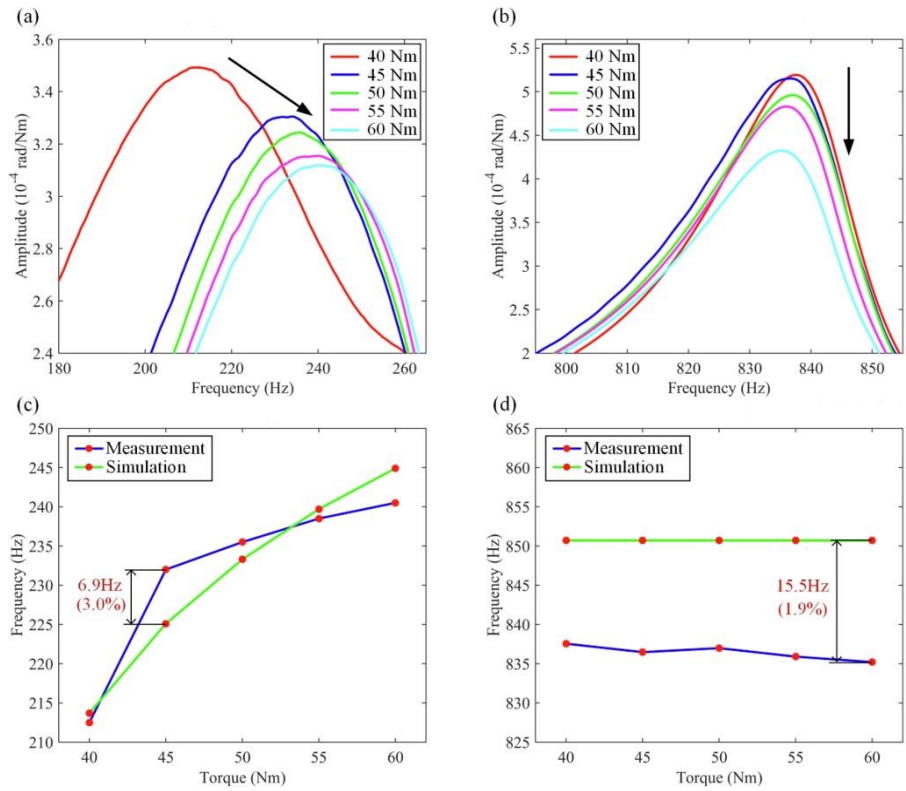

Figure 9 Influences of lead screw pre-stretching on system natural frequency. (a) The 1st order origin point amplitude-frequency response of the lead screw. (b) The 2nd order origin point amplitude-frequency response of the lead screw. (c) The relationship between the 1st order natural frequency and the torque. (d) The relationship between the 2nd order natural frequency and the torque [22]

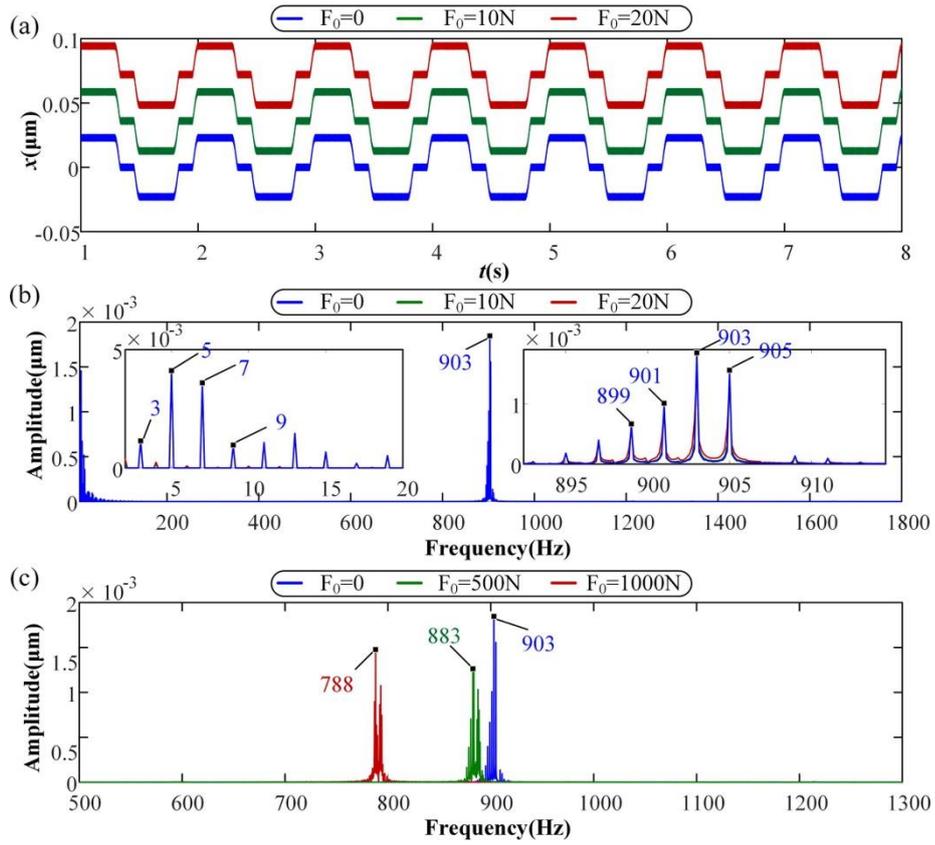

Figure 10 Time and frequency domains of axial vibration with varied external loads (a) Time histories when external loads $F_0$ = 0, 10 N and 20 N, (b) Frequency spectra when external loads $F_0$ = 0, 10 N and 20 N, (c) frequency spectra when external loads $F_0$ = 0, 500 N and 1000 N [63]

The vibration could be reduced by decreasing the small ball dimension and the phase difference between the ball groups[63]. Additionally, the nominal contact angles of the screw-nut are a critical parameter affecting system performance [58]. Larger nominal contact angles increase the fundamental resonance frequency and decrease the primary harmonic amplitude [30].

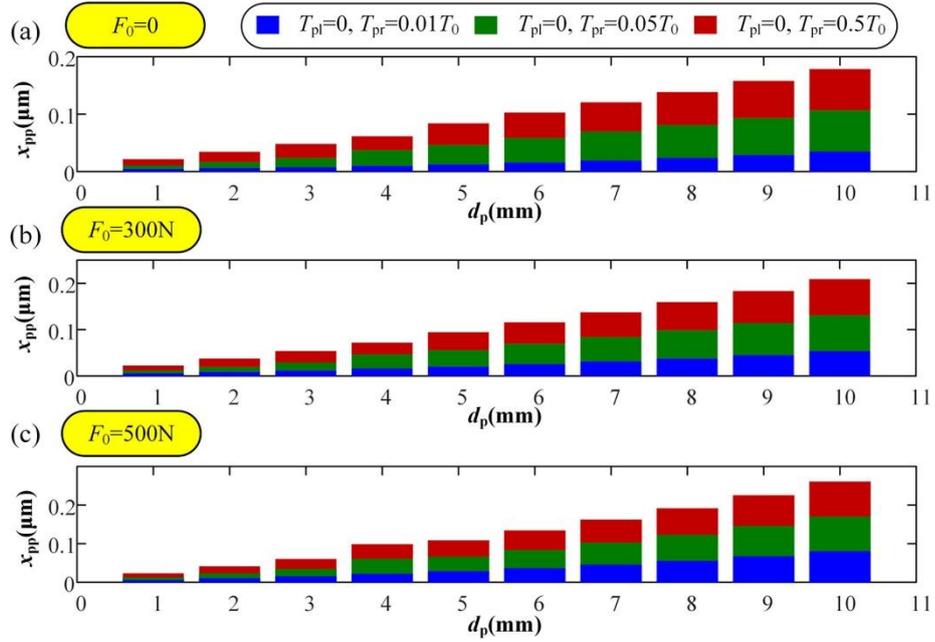

Figure 11 Peak to peak values of axial vibration displacement with varied ball diameters; (a) peak to peak value with varied ball diameters when external load $F_0$ = 0, (b) peak to peak value with varied ball diameters when external load $F_0$ = 300 N, (c) peak to peak value with varied ball diameters when external load $F_0$ = 500 N. ($T_{pl}$ and $T_{pr}$, the lead phases of ball in the left and right nuts, respectively; $T_0$, ball passage period; $d_p$ pitch circle diameter of screw shaft) [63]

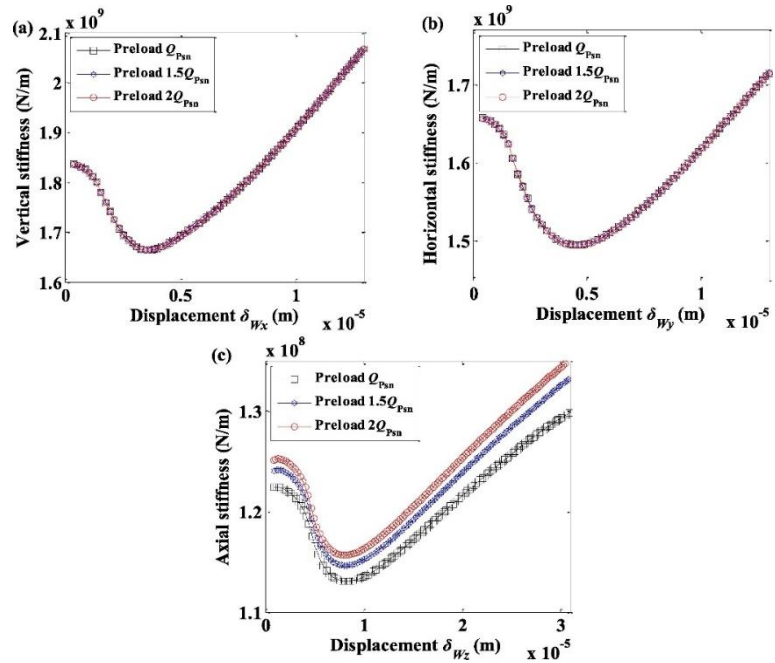

Figure 12 Stiffness curves with different preloads of the ball screw. (a) vertical stiffness; (b) horizontal stiffness;(c) axial stiffness. [58]

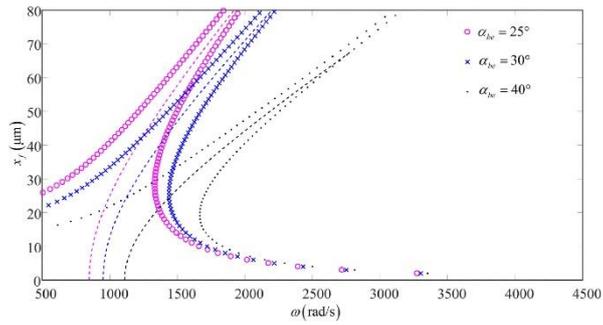

Figure 13 Fundamental response curves in various nominal contact angles of screw-nut joints [30]

### 2.1.3 Linear guide

The application of rolling linear guides in precision machine tools has several advantages; however, their unstable and flexible dynamics due to ball and groove contact result in strong nonlinearity and poor chattering stability[16,64–66]. Experimental results by Hadraba et al. [67] showed that the frequency response of rolling linear guides significantly changes under specific cutting forces and vertical loads. They also found that increasing the preload of linear guides enhances the feed system's stiffness [58].

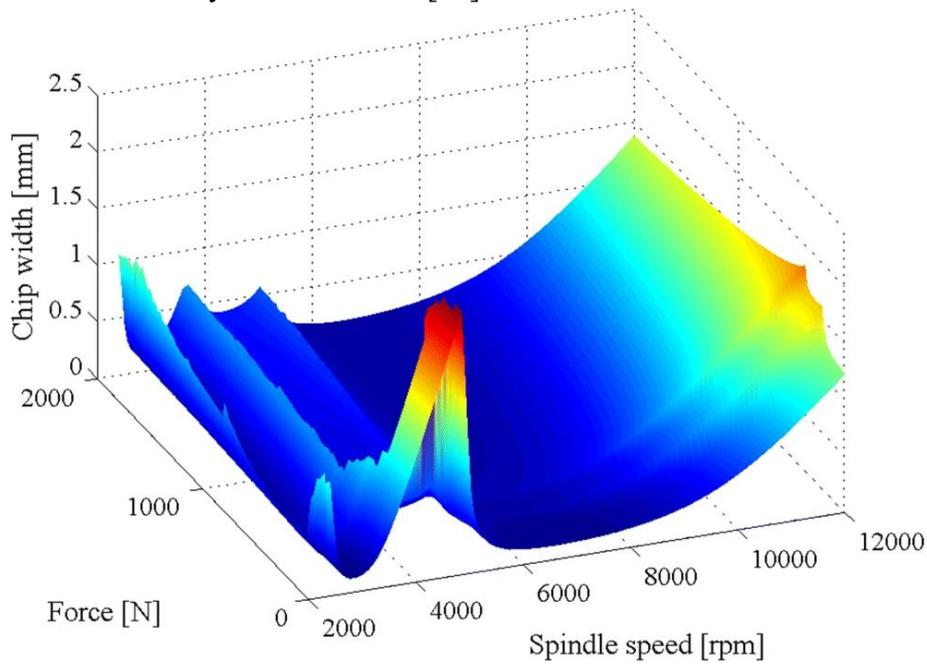

Figure 14 Estimated linearised chatter stability lobe diagram of LBG structure as function of static load [67]

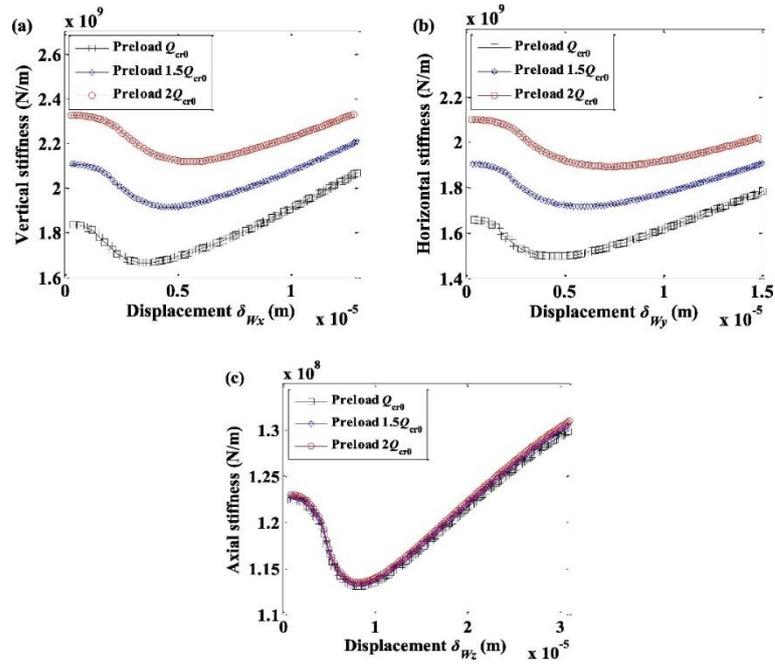

Figure 15 Stiffness curves with different preloads of the guide rail. (a) vertical stiffness; (b) horizontal stiffness;(c) axial stiffness.-[58]

Through the reciprocating sliding experiments of the sliding guide under similar working conditions, Yue et al. [68] showed that the suitable surface weave can reduce guide friction, alleviating the guide stick-slip phenomenon[31,69,70].

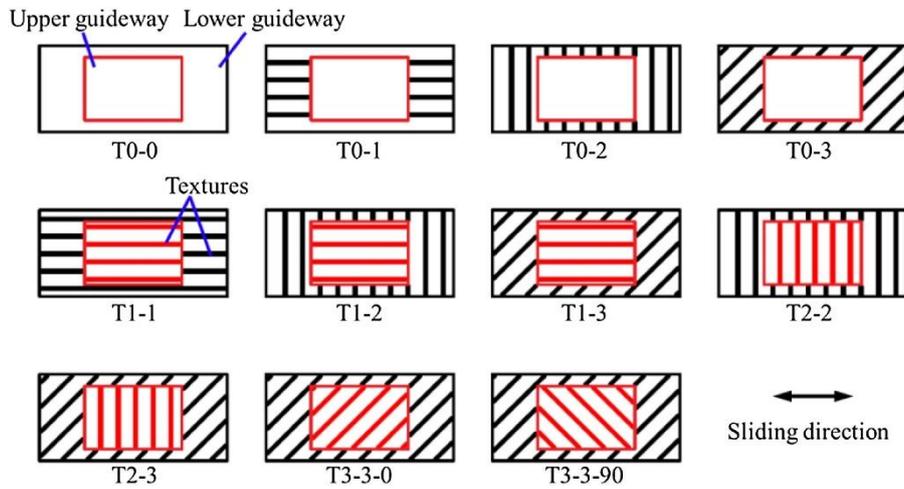

Figure 16 Nomenclature used for the texture combinations [68]

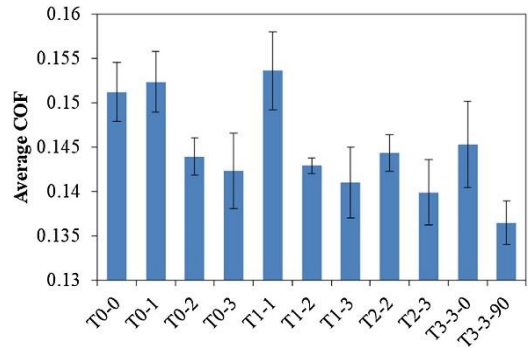

Figure 17 Average coefficients of friction (COF) of different textured pairs [68]

The guide profile curve directly affected the straightness and guide motion angular error. Due to the randomness of the manufacturing and assembly process, a deviation from the traditional method of using a sine curve to describe the guide profile curve occurs[71,72]. Therefore, Niu et al. [23] proposed a method for static analysis and motion error prediction of the guide assembly under the uncertainty of the thread friction coefficient. The prediction accuracy was above 81%.

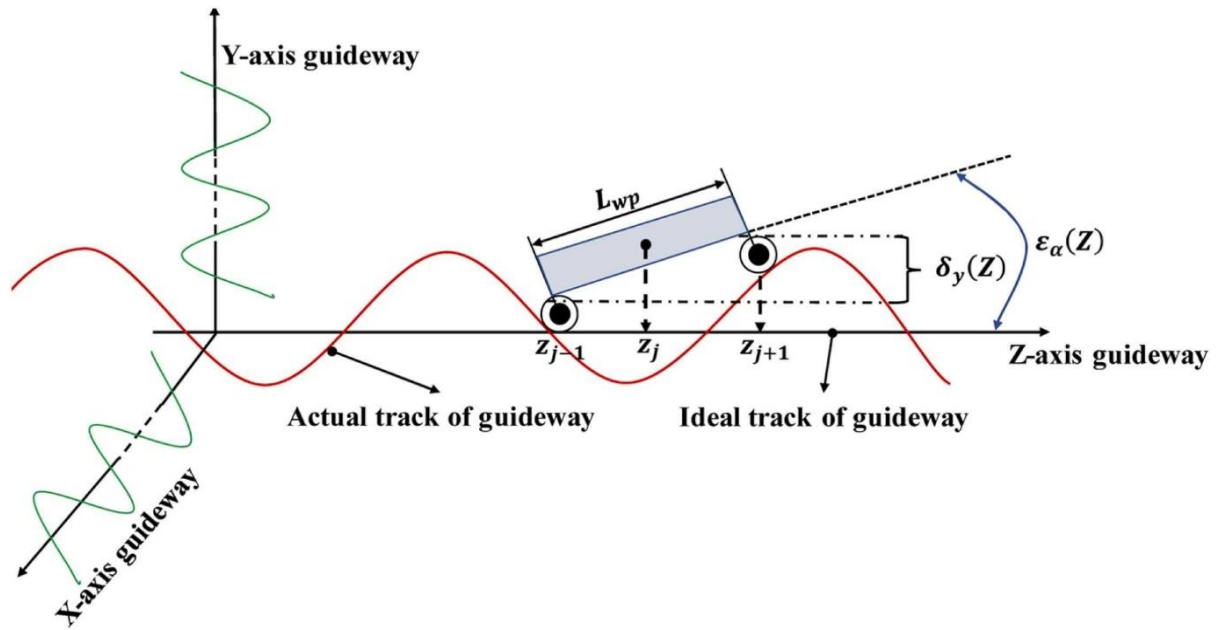

Figure 18 Schematic view of the working platform's movement along the guideway [23]

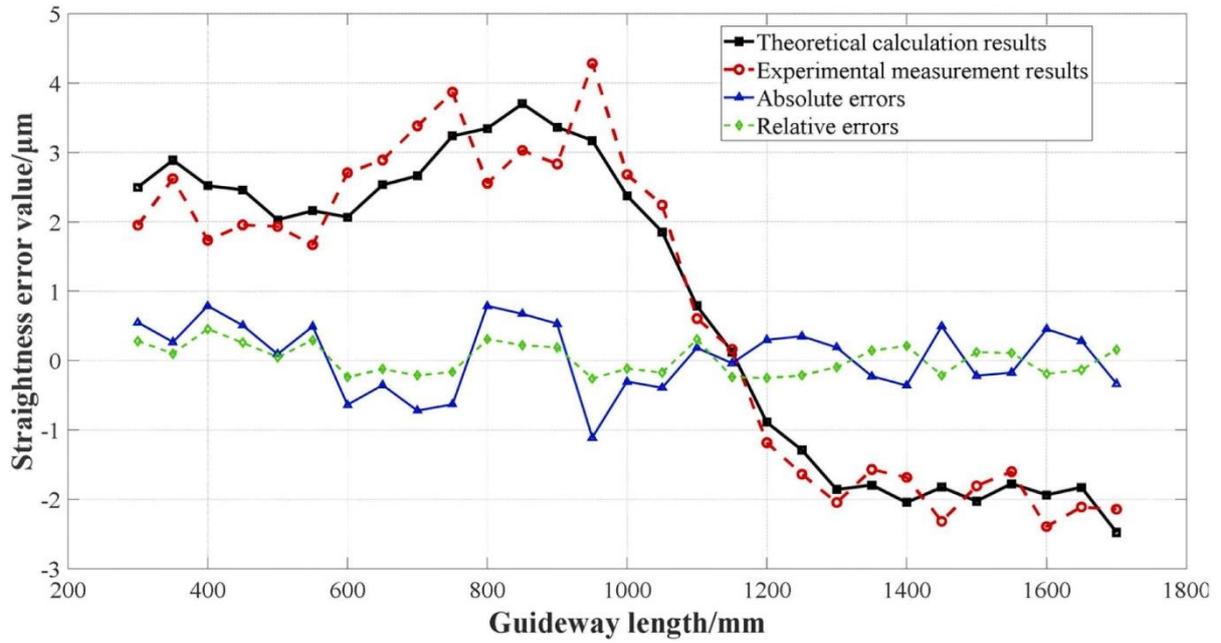

Figure 19 Schematic view of the compared results for straightness error [23]

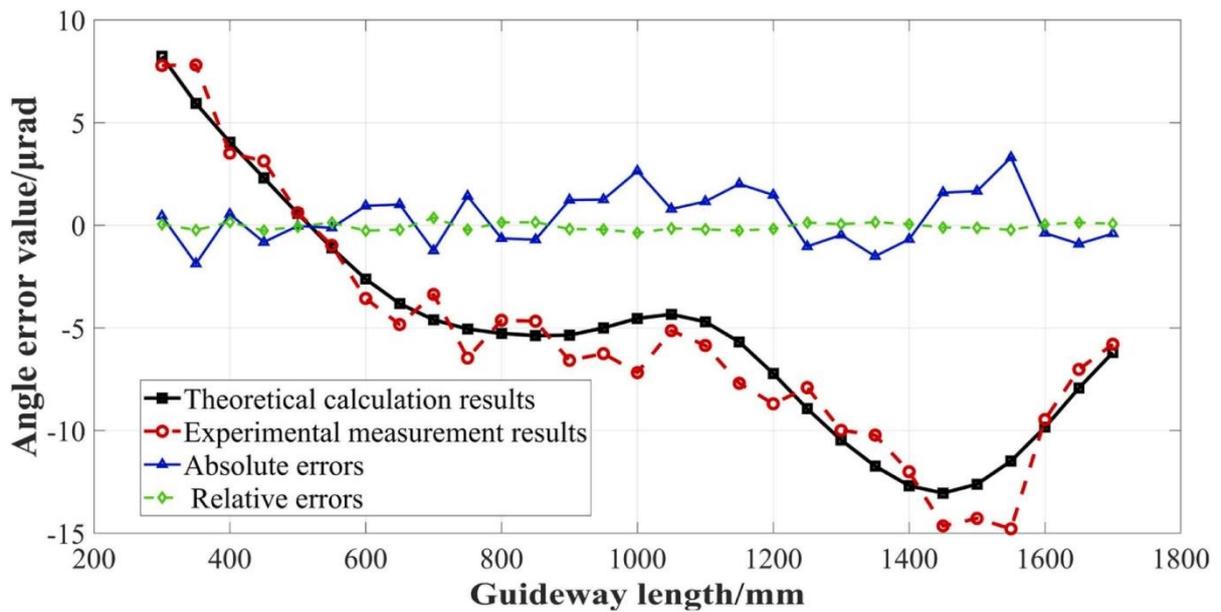

Figure 20 Schematic view of the compared results for angle error [23]

### 2.1.4 Bearing

Due to centrifugal force and external load, bearings produce elastic deformation, seriously affecting their vibration conditions[73–75]. Aiming to conduct an in-depth analysis of their vibration characteristics, Liu et al. [76] analyzed the effects of external load and rotational speed. The results have shown that the elastic deformation of rollers and raceways significantly affects their vibration characteristics.

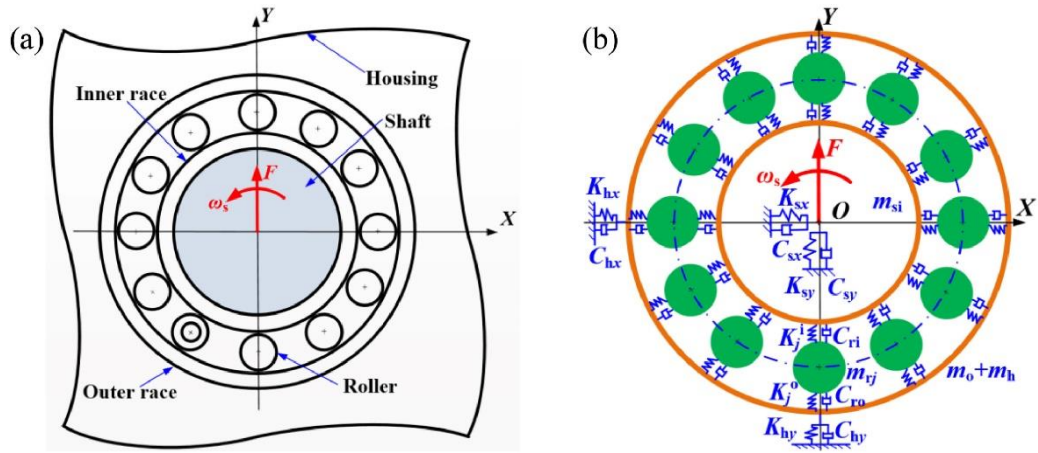

Figure 21 Schematics of (a) a CRB system and (b) a dynamic model for the CRB system [76] (Note: CRB - cylindrical roller bearings)

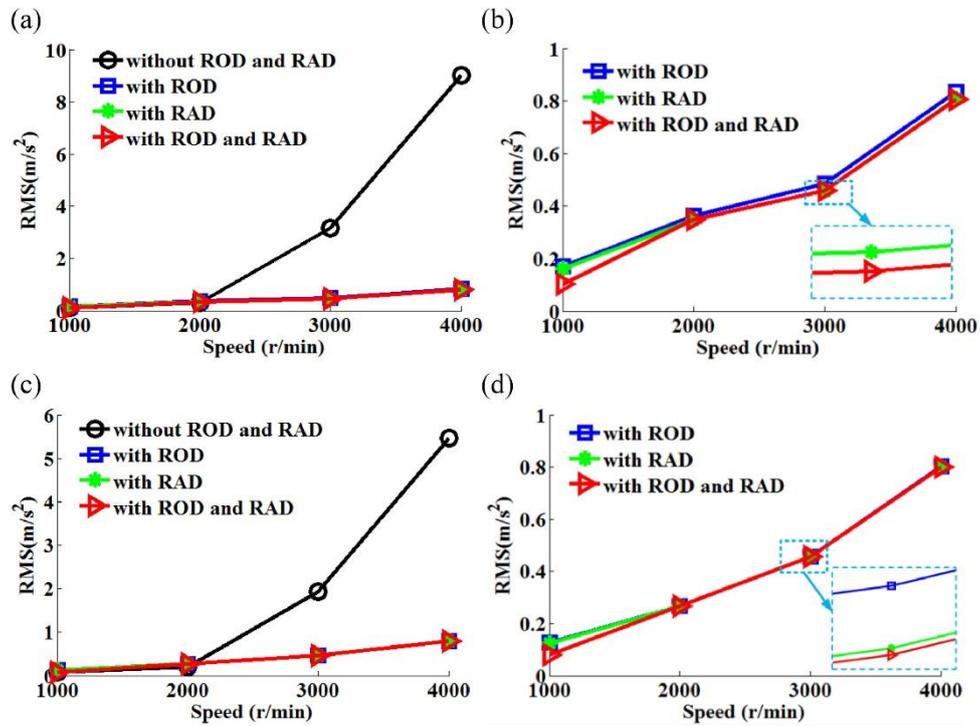

Figure 22 Influence of the inner race speed on the RMS value of the accelerations of the inner and outer races for the rigid and flexible cases. (a) Inner race, (b) enlarge view of (a), (c) outer race, and (d) enlarge view of (c). (Note: ROD denote the roller deformation; and RAD denote the race deformation) [76]

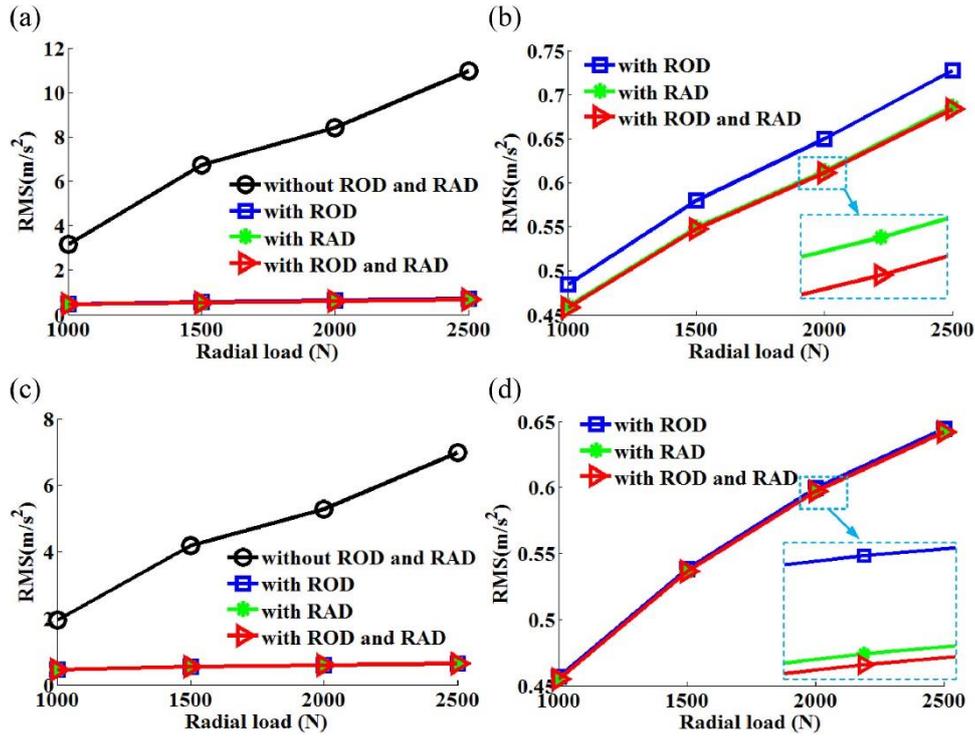

Figure 23 Influence of the radial load on the RMS value of the accelerations of the inner and outer races for the rigid and flexible cases. (a) Inner race, (b) enlarge view of (a), (c) outer race, and (d) enlarge view of (c). (Note: ROD denote the roller deformation; and RAD denote the race deformation) [76]

The load distribution and radial displacement of rolling bearings are directly related to their accuracy and lifetime. Aschenbrenner et al. [77] investigated the complex dependence between the load distribution and radial displacement of a bearing, as well as its influence on geometric errors.Next, Yakout et al. [78] studied the effect of internal radial clearance of rolling bearings on their damping characteristics, natural frequency. The damping ratio decreased with the increase in clearance; the internal radial clearance of rolling bearings significantly affected the natural frequenct.

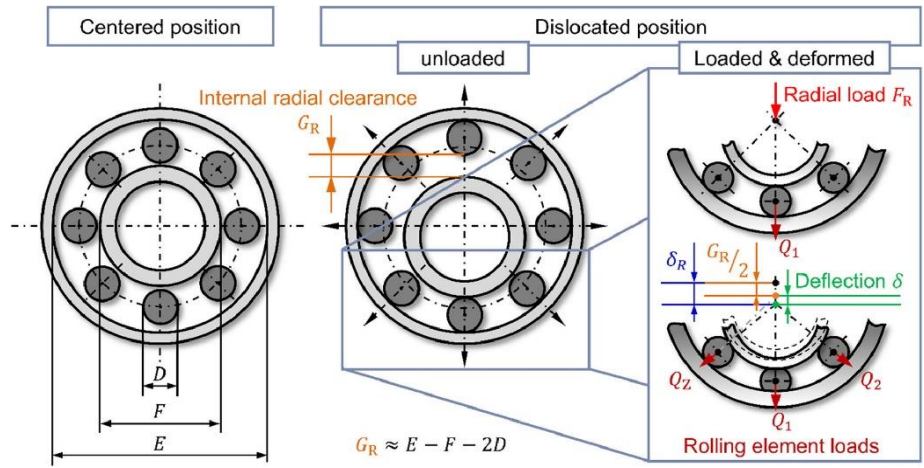

Figure 24 Schematic comparison of the centered and dislocated positions of a radially loaded rolling bearing [77]

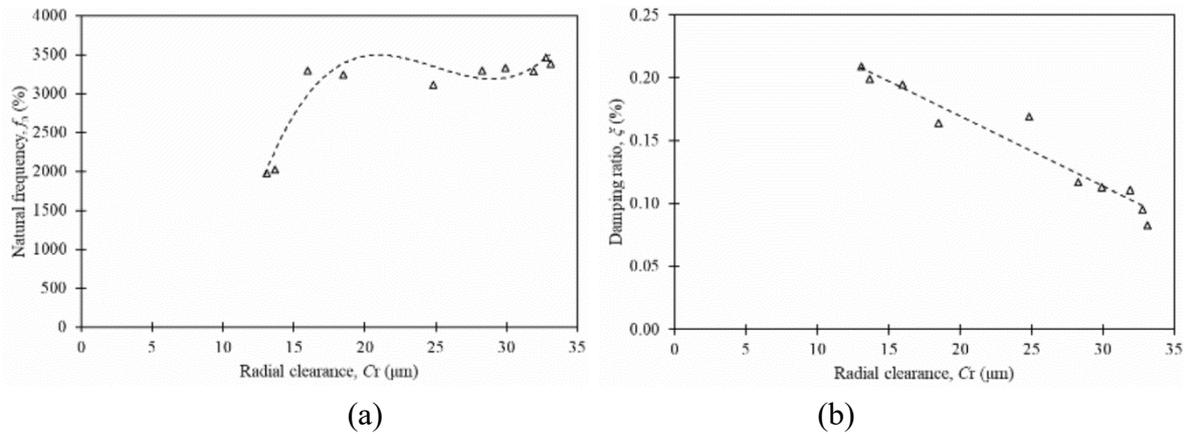

(a)            (b)

Figure 25 Relationship between the radial clearance and (a) natural frequency (b) damping ratio of the bearings tested [78]

The nominal contact angles of bearing joints also significantly impact the feed system's performance, with larger angles increasing the fundamental resonance frequency and reducing the primary harmonic amplitude [30,58].

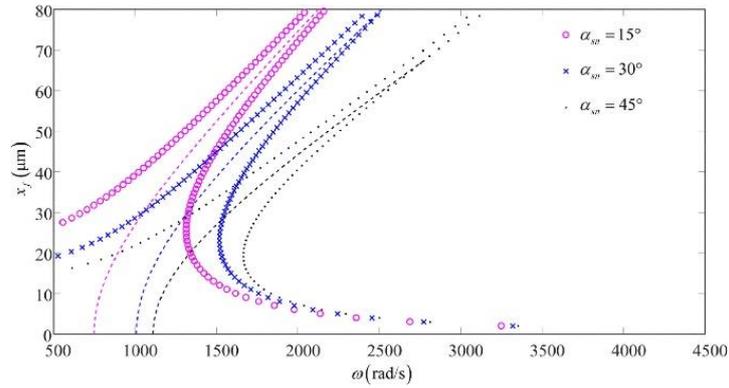

Figure 26 Fundamental response curves in various nominal contact angles of bearing joints [37]

## 2.2 Motor performance optimization

The motor is the power source of the CNC machine tool feed system; in the motor optimization process, the goal is generally to reduce the rotational inertia of the rotor and increase the torque, high torque accuracy, and high-speed performance [79–85].

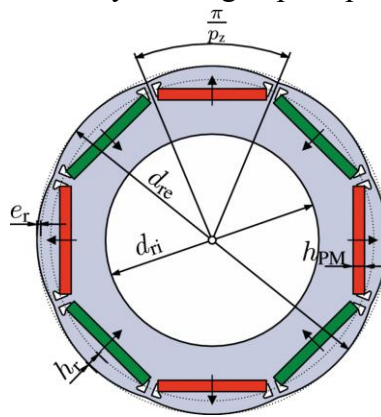

Figure 27 An example for a parametrized rotor geometry [86]

Bramerdorfer et al. [86] summarized the recent advances and new trends in motor design optimization, concluding that the optimal motor design should depend on the specific scenario – there is no universal global optimal design solution. Zhu et al. [87] proposed a multimode design method for motors operating in multiple working conditions. The method is based on the driving conditions of electric vehicles and identifies five typical operating conditions and five corresponding driving modes. These modes are needed for the multimode design of motors, ensuring good performance in multiple driving modes.

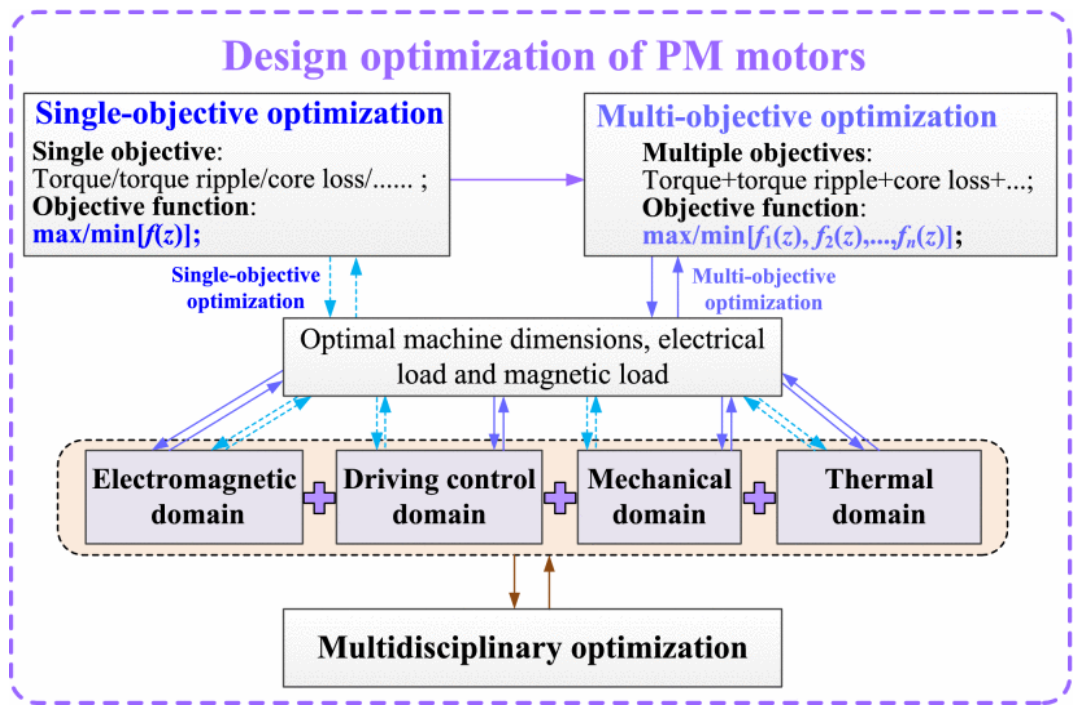

Figure 28 Classification of PM motors design optimization [87]

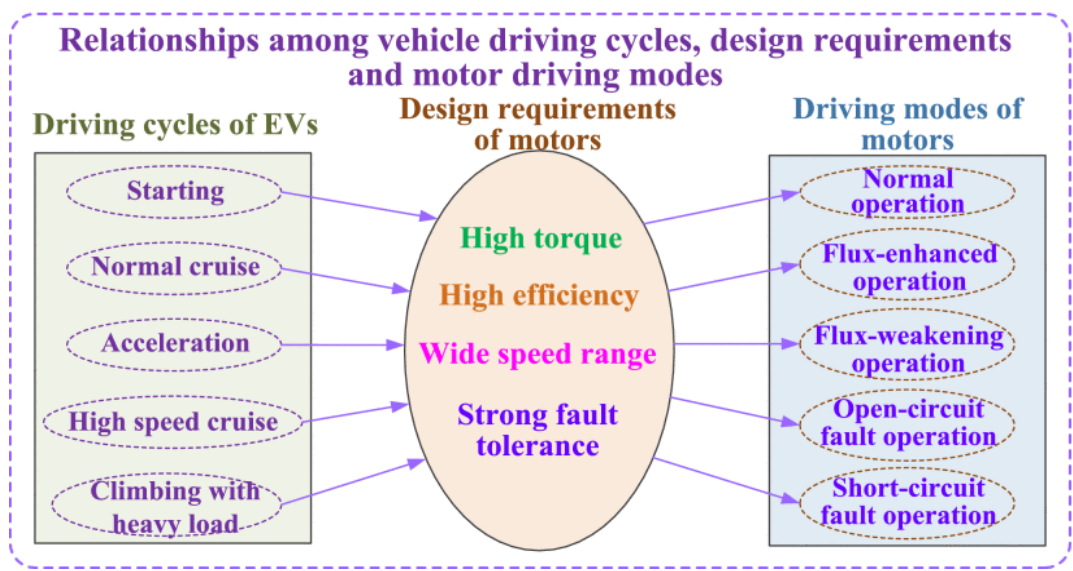

Figure 29 Relationship among vehicle driving cycles, design requirements, and motor driving modes [87]

Lei et al. [88] used the space reduction optimization (SRO) strategy to optimize three robust design methods for permanent magnet motors, considering material diversity and manufacturing tolerances. Their results showed that the proposed SRO strategy significantly improves the optimization effectiveness and computational efficiency of these traditional robust design methods. Kumar et al. [89] optimized the motor design by changing the span of stator teeth to change their tooth shape, thus reducing the cogging torque of the brushless DC motor. Based on this idea, the scheme was found to reduce the peak-to-peak cogging torque and reduce the vibration and noise during the motor operation. Wang et al. [90] performed a rotor design optimization of a permanent magnet synchronous motor using a multi-objective optimization algorithm. The algorithm was used to solve the rotor design parameters and maximize the output torque under three different load conditions – no load, rated load, and the maximum speed load. The optimized permanent magnet synchronous motor has high torque and high-speed performance and is expected to achieve wide speed range machining. This included both low-speed heavy-load and high-speed cutting without replacing the motor or machine tool.

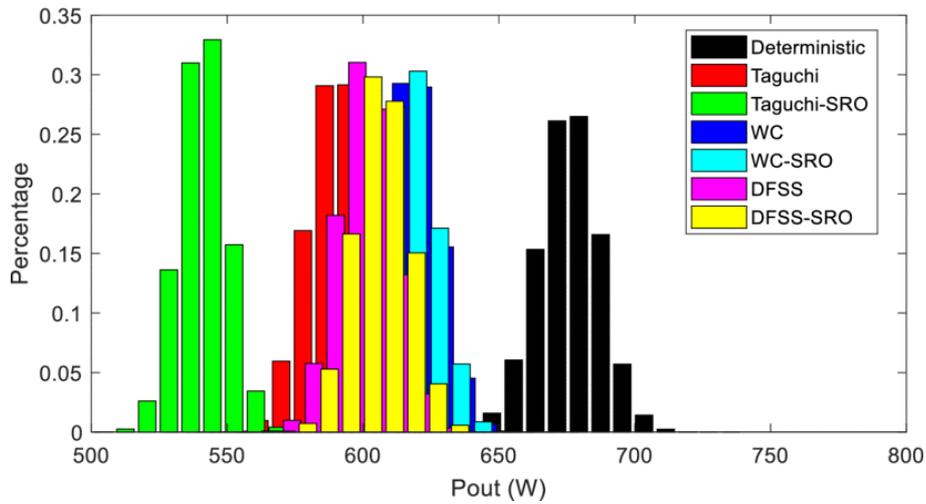

Figure 30 Distributions of output power for different methods [88]

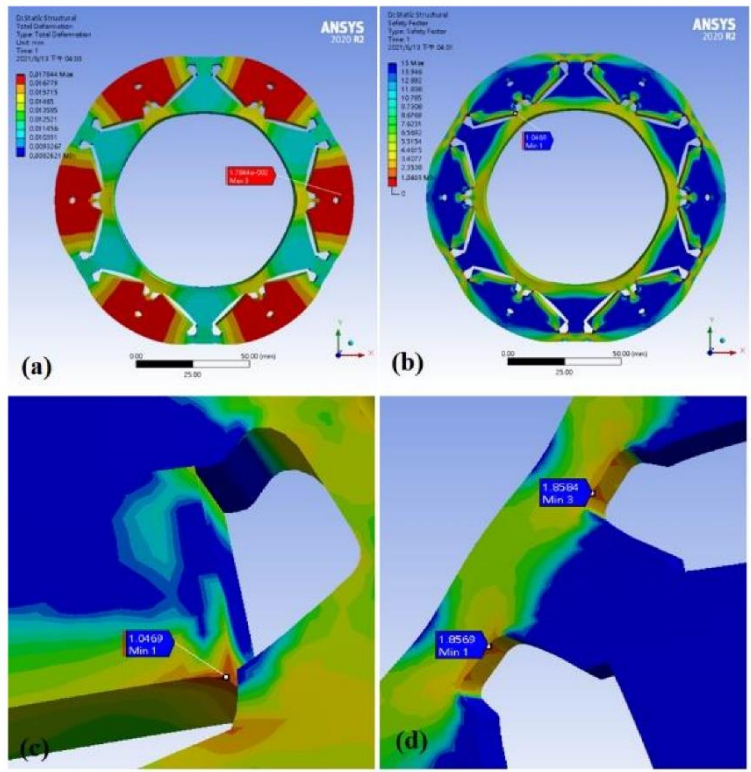

Figure 31 Simulation results for: (a) rotor deformation, (b) safety factor of rotor assembly, (c) minimum safety factor in permanent magnet and (d) minimum safety factor in rotor [90]

## 2.3 Control algorithm optimization

Under certain system dynamic performance requirements, the control algorithm can be optimized after the dynamics matching between the motor and mechanical entities is completed [55,91–93]. The control system was optimized using parameter optimization, trajectory prediction algorithm optimization, and by adding compensation modules to improve the system dynamic performance. The control algorithm optimization helps provide the best performance of the electromechanical system[94–100].

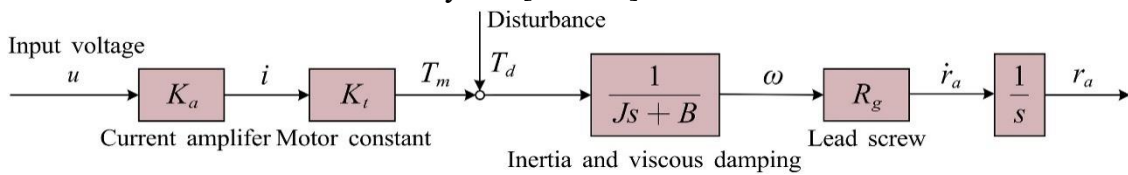

Figure 32 Block diagram of the feed drive system [101]

To improve the contour motion control effect without affecting the original closed-loop controller structure of the precision multiaxis systems, Wang et al. [102] proposed a modular design method for trajectory compensation. The proposed method was based on a precise task coordinate system, a compensation module independent of the existing closed-loop control system. For this reason, it is an important reference for optimizing the control performance of existing CNC systems molded in packages.

Du et al. [101] proposed a contour error estimation method based on the third-order meshing helix. The proposed method optimally adjusts the speed tracking and contour error coefficients to improve the contour control accuracy. Yang et al. [103] developed a nearest point projection curvature circle iterative (NPP-CCI) algorithm. The experimental results show that the NNP-CCI algorithm has higher contour error estimation accuracy than the traditional unified curvature circle iterative algorithm.

The study by Xie et al. [36] first disclosed the tracking error generation mechanism in parallel machine tools. In other words, the tracking error of the driving body includes the error caused by time-varying load and the error caused by the input signal. A comprehensive control method combining real-time dynamic feedforward and feedrate scheduling optimization was proposed for these two tracking error segments, effectively improving the machining efficiency and quality. Kim et al. [104] proposed a robust real-time tool trajectory correction method based on a state estimator; since the offline tool trajectory correction method did not consider the error due to model uncertainty, a real-time contour error estimation method using interpolated data was proposed. The method effectiveness was verified through experiments, by comparison to the traditional offline tool trajectory correction methods. Lyu et al. [105] conducted a study on the lunar trajectory showing that the servo outer loop error also affects the trajectory tracking error.

## 2.4 Motion process optimization

Following the continuous optimization of the feedrate scheduling and interpolation algorithm by the researchers, the processing quality and processing efficiency displayed very good results. Good feedrate planning and interpolation algorithm can reduce abrupt speed, acceleration, and jerk changes, improve machining accuracy and quality[106–108]. All these characteristics are important when achieving the high speed and high precision machining of the CNC machine tool feed system[109].

### 2.4.1 Feedrate planning

Feedrate planning is the core of the feed motion system and is critical when improving machining efficiency, accuracy, and subsequent quality [110,111]. Liang et al. [112] proposed a general feed speed optimization scheduling method considering geometric and drive constraints. They represented the speed curve using B-spline curves and used genetic algorithms to solve the constructed feed speed optimization model. This method can efficiently and robustly generate globally time-optimal feed speeds.Lin et al. [113] proposed a fuzzy feedrate planning method based on curvature and curvature variation. The experimental results showed that for the same cutting time, the method could improve the cutting accuracy by 41.8%; on the other hand, the cutting time was reduced by 50.8% for the constant cutting accuracy. Sun et al. [114] proposed a computational optimization algorithm for feedrate planning that satisfies both machine drive condition constraints and process constraints; the feedrate was used as the main optimization variable. The analytic linearization of machining-related constraints based on machine axis velocity, axis acceleration, and axis displacement was carried out. The experimental results have shown that the method can achieve efficient feedrate planning that satisfies both the physical and limit constraints of the machine. A review by Sun et al. [115] provided an overview of the current status. The conclusion is that feed speed planning methods should evolve to consider dynamic characteristics to further enhance the potential of machine tools.

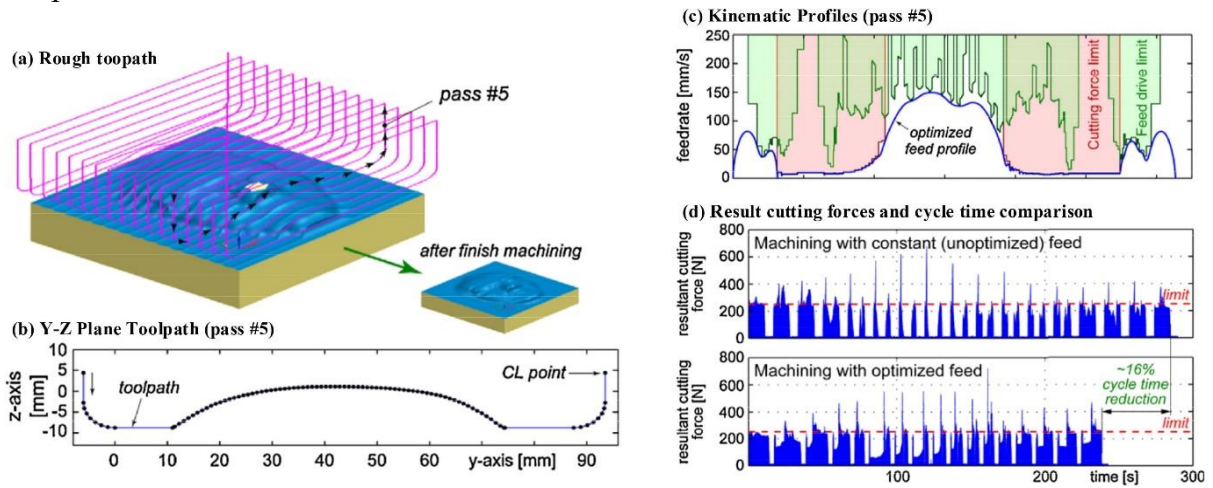

Figure 33 Feedrate optimization for free-form milling considering constraints from the feed drive system and process mechanics [115,116]

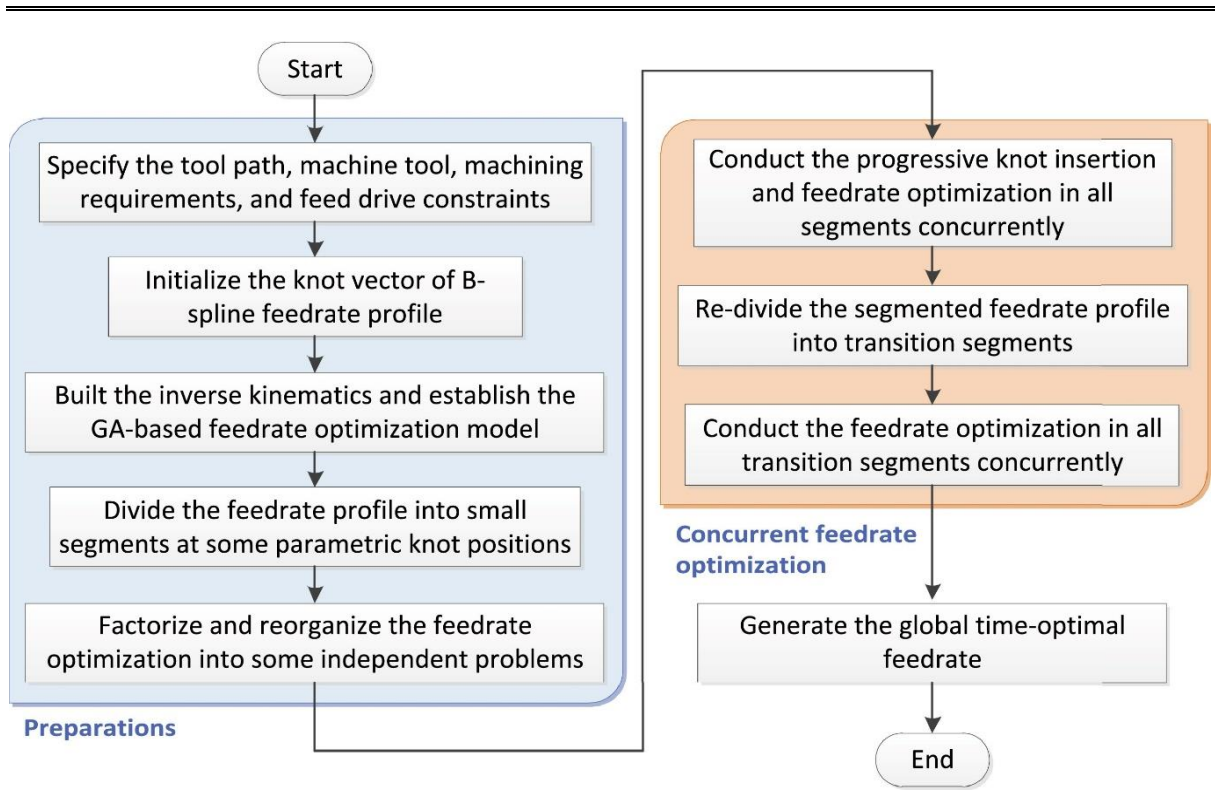

Figure 34 Flowchart of the optimization-based feedrate scheduling method [112]

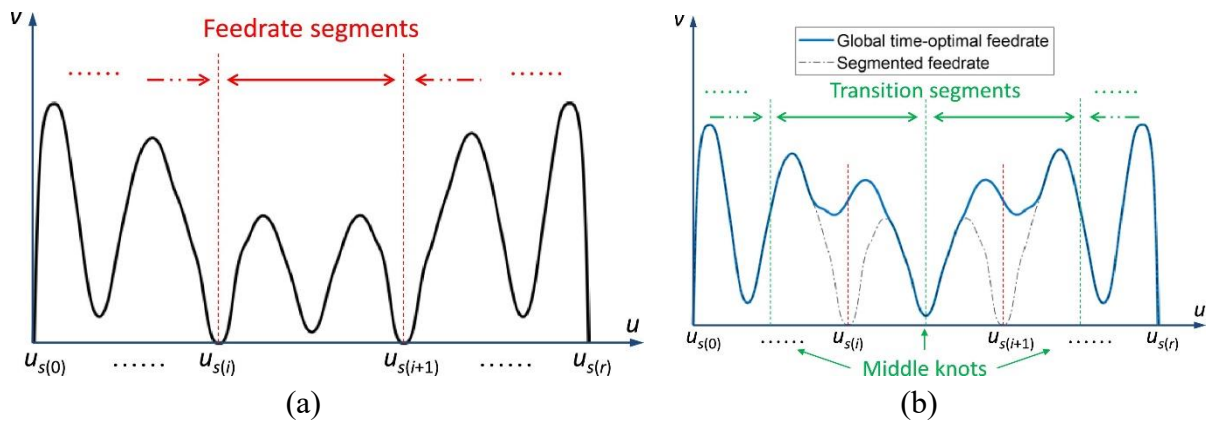

Figure 35 Concurrent feedrate optimization in (a) the divided segments (b) transition segments [35,112]

## 2.4.2 Interpolation algorithm

The vibration characteristics of the feed system were improved by continuously optimizing the interpolation algorithm to maximize the trajectory smoothness. Smooth, reasonable, and high-quality CNC instructions are prerequisites for high-speed and high-precision performance. The tool trajectory in high-speed machining comprises discontinuous G01 line segments generated through computer assistance[117,118]. The discontinuity of tool motion leads to low machining efficiency of CNC machine tools. Various interpolation algorithms were studied to achieve high-speed continuous tool motion ; however, optimizing the smoothed trajectory in a real-time system is difficult. For this reason, Li et al. [119] proposed a direct trajectory smoothing method based on neural networks. Although this method uses machine learning methods, it does not require iterative learning and repetitive computation. These characteristics ensure the real-time performance, simple structure, low computational cost, and straightforward application to real-time CNC systems.

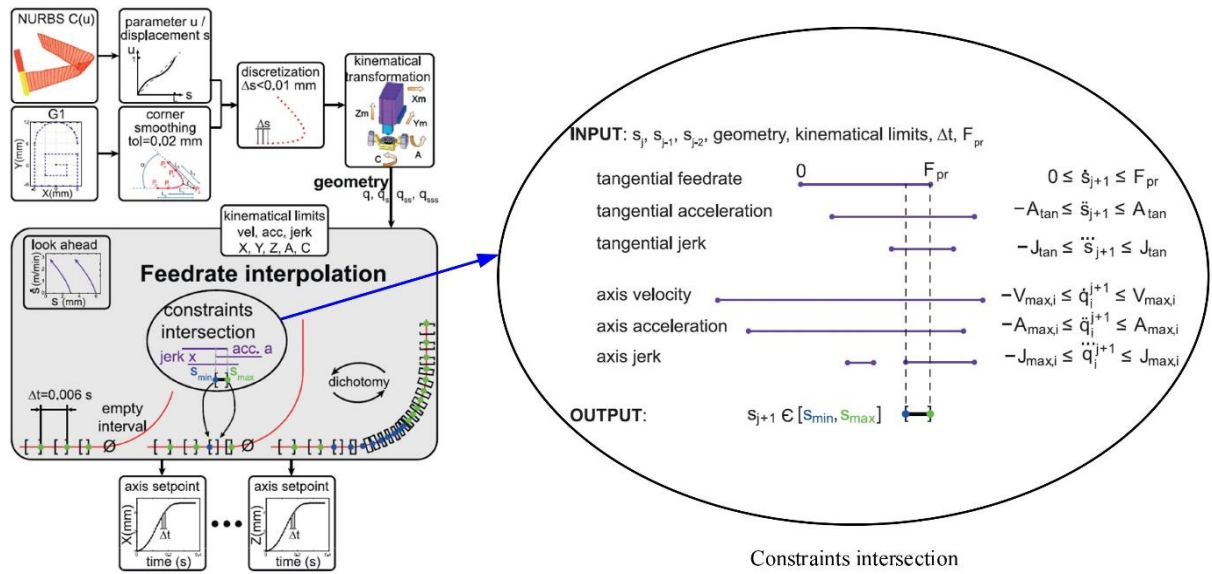

Figure 36 Velocity profile optimization based dichotomy strategy and constraint intersection [115,120]

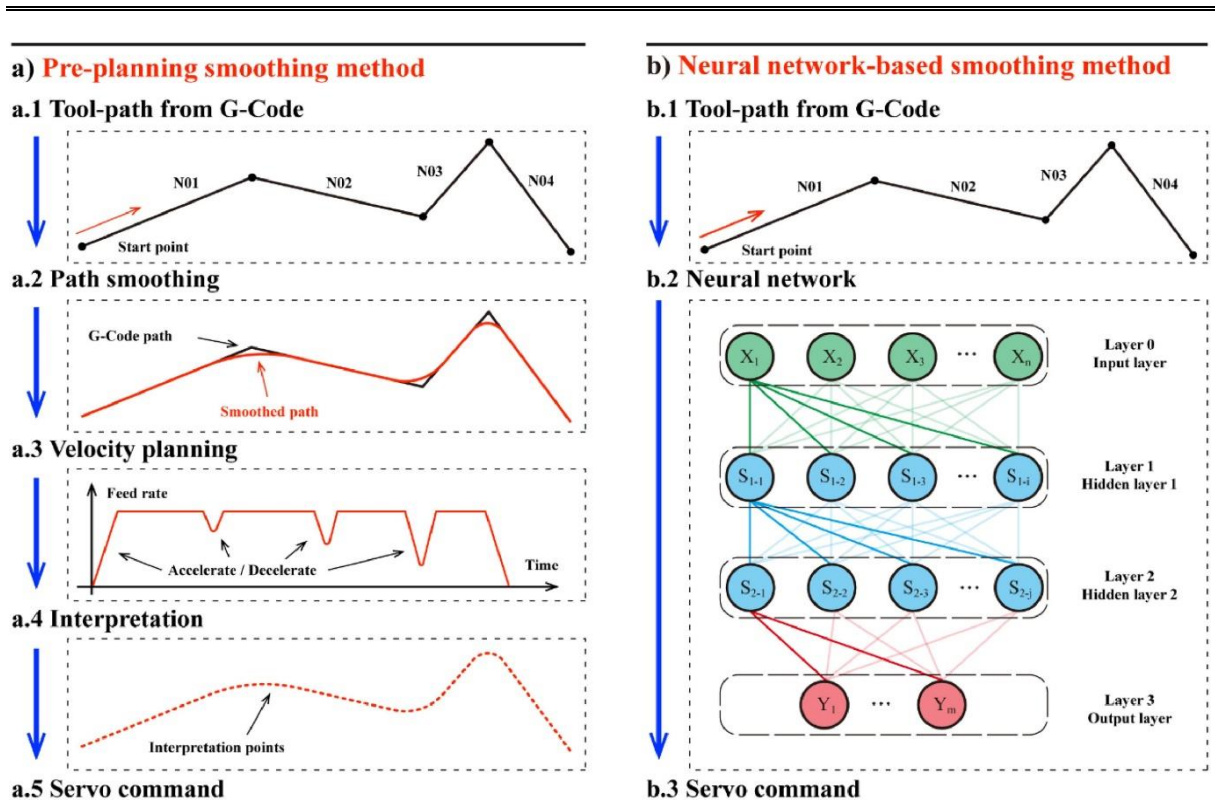

Figure 37　Comparison of the pre-planning-based smoothing method and the proposed neural network-based method [119]

Zhang et al. [121–123] proposed a trajectory smoothing algorithm for generating G4 continuous tool trajectory by five B-sample curves. The authors aimed to create the tool feed with smoothly changing acceleration, improving it based on G4 continuous tool trajectory and acceleration-smoothed feedrate planning. The algorithm was compared via point-to-point motion, fixed feedrate strategy, and G2 and G3 methods, respectively, verifying the feasibility and effectiveness of the method. This enabled avoiding unnecessary feed setbacks and inertial shocks, balancing the time optimality and motion performance. Zhang et al. [124] proposed an acceleration smoothing algorithm based on fluctuating finite acceleration profiles, reducing the overall machining time by approx. 6-7%. The curvature-smoothed motion trajectories, velocity profiles, and acceleration profiles were obtained.

# 3　Study of the coupling mechanism between sub-systems

The CNC machine tool feed system comprises several sub-structural components and functional units. There are also complex coupling relationships between the sub-systems [38,39]. Since the mechanical system, motor, control system, and motion process collectively influence system performance, optimizing the design of individual subsystems may not achieve the best overall performance of the feed system.

For example, researchers have lightened the worktable through structural optimization and material filling to improve its natural frequency and enhance the high-speed and high-precision performance of the CNC machine tool feed system [27–29,62]. However, Zhang et al. [37] found that the load inertia ratio of the feed system should be kept within a certain range to maintain good dynamic performance under high-speed and high-acceleration conditions, as shown in Figure 38. This means that excessive lightweight design of the worktable should be avoided, and its matching relationship with the motor should be considered.

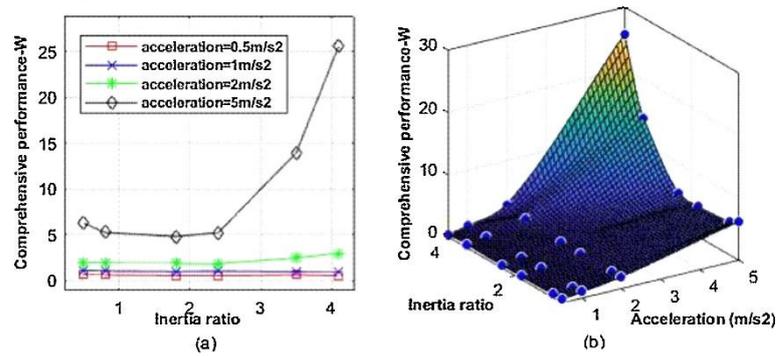

Figure 38 System performance changes corresponding to different inertia ratios and accelerations

Additionally, there are complex interactions such as inertia matching and electromagnetic excitation in electromechanical systems. For example, adjusting control system parameters also affects the performance of the electromechanical system [40]. Changes in the motion process affect the frictional force of the electromechanical system, altering interface characteristics and consequently the system's stiffness and damping properties [41–43,125–128]. The motion process also affects inertial forces, changing the system's vibration and deformation, thereby altering the electromechanical system's dynamic characteristics [44].

Given the complex coupling interactions among subsystems and their collective impact on dynamic performance, comprehensive research on these interactions is essential. Such research can guide the optimal design of CNC machine tools by considering these interactions, thereby maximizing dynamic performance.

Currently, research on the coupling mechanisms of CNC machine tool feed system subsystems mainly focuses on the tuning of control parameters considering mechanical characteristics [40], inertia matching between the motor and mechanical load [37], and the impact of motion processes on mechanical system interfaces [41–43,125–128]. However, there is a lack of comprehensive and systematic research encompassing the interactions of all subsystems.

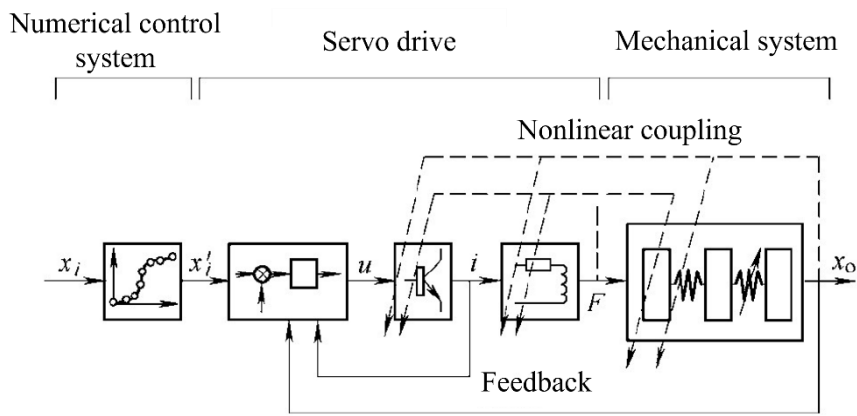

Figure 39 Electromechanical coupling of the feed system[39]

## 3.1 Coupling mechanisms of mechatronic systems

Electromechanical control systems are tightly coupled, with complex mutual influences between the motor and mechanical structure, and adjustments to control parameters can alter the performance of the entire electromechanical system.

Weck et al. [40] investigated the coupled oscillation effect between the control system and the mechanical structure, calculating the adjustment range of the controller parameters. In researching the value range of control parameters, the mechanical structure's characteristics should be considered. They simplified the machine tool feed system to a simple single mass vibrator, the Nyquist criterion can be used to provide stability conditions for the speed and position control loops, considering driver limitations (maximum current and voltage). As shown in Figure 40, these factors yield the tuning range for controller parameters ($Kp$: Gain of position controller, $Kv$: Gain of speed controller, $Tp$: Reset time of speed controller).

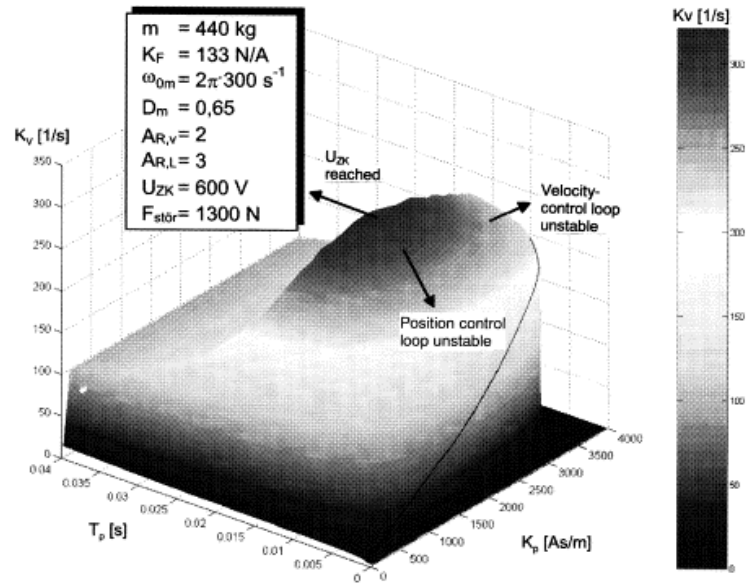

Figure 40 Setting range for controller factors with respect to system stability and physical limits[40]

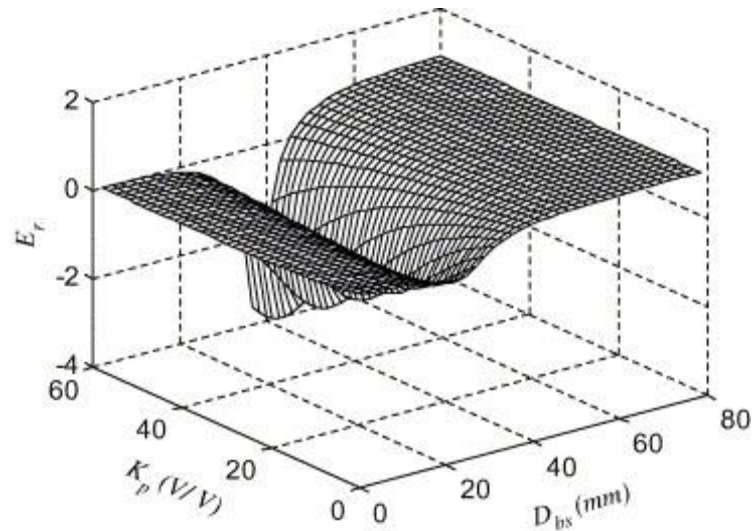

Figure 41 Effects of mechanical–control interaction on the contour error ratio [129]

Szolc et al. [130] studied the coupling characteristics between a rotating mechanical system and a motor, establishing an analytical computational model. As shown in Figure 42, asynchronous motors generate higher electromagnetic damping within a small torsional excitation frequency range. Electromagnetic damping more effectively dissipates torsional vibration energy compared to mechanical damping, suppressing the resonance effects of the driven mechanical system. The study enabled the designer to select the optimally matched motor according to the characteristics of the driven mechanical part.

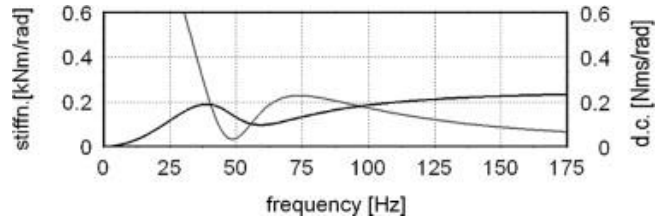

Figure 42 Electromagnetic stiffness (black line) and damping coefficient (gray line) [130]

In the ball screw drive system, the transmission path of torque transfer from the motor to the table displays a buffering effect on the electromechanical coupling characteristics of the motor and the mechanical part. On the other hand, the linear motor has no buffering effect, and the corresponding coupling is rather apparent [131–135].

Yang et al. carried out multiple studies on the electromechanical coupling effects of the linear motor feeding system. They proposed an integrated modeling and analysis method for the electromechanical coupling characteristics of linear motor direct drive feeding systems [38]. Four electromechanical coupling phenomena were systematically analyzed and compared. Coupling caused by thrust harmonics is a major factor affecting displacement fluctuations. In high-speed, high-precision applications, coupling due to air gap fluctuations becomes prominent. Encoder errors, determined by encoder installation and position, significantly impact displacement fluctuations. The coupling effect of outside disturbances depends on the disturbance characteristics.

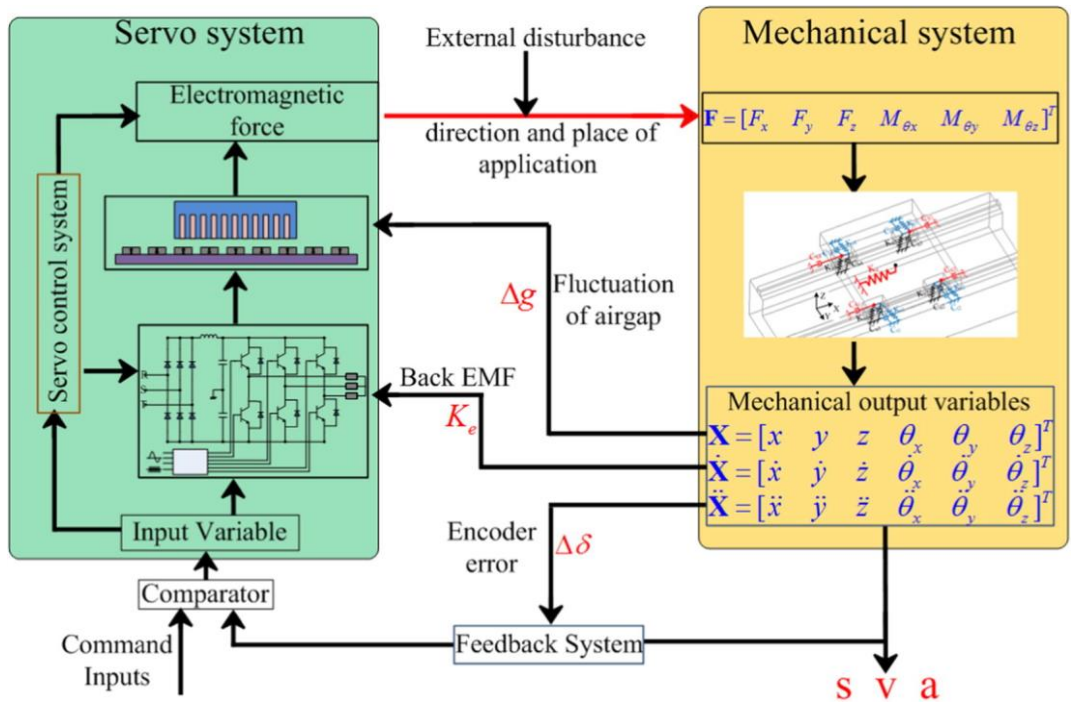

Figure 43 The coupled process of the electromechanical coupling[38]

Subsequently, the same research group analyzed the effect of motor thrust harmonics on electromechanical coupling characteristics [136–138]. They found that mechanical vibration and motor thrust characteristics affect each other by coupling, resulting in poor dynamic accuracy of the feeding system. Additionally, the thrust harmonic characteristics varied with the moving process and load. Further, it was also found that mechanical vibration couples with magnetic field harmonics and drives the circuit harmonics to generate a large number of new thrust harmonics. Such behavior diminishes the dynamic accuracy of the motion system [139,140].

The above-presented studies show that the instability of the motor output torque and the mechanical system vibration couple, resulting in complex interactions. Control parameter research has also explored the mechanisms by which control parameters affect electromechanical systems. However, their effect on the electromechanical coupling law under different conditions of motion processes requires further study.

## 3.2 Coupling mechanisms between motion planning and mechanical properties

The excessive friction caused by changes in the moving process can affect the joint surface characteristics of the ball screw feed system components, in turn affecting the dynamic performance of the whole system [41–43,125–128], as shown in Figure 44.

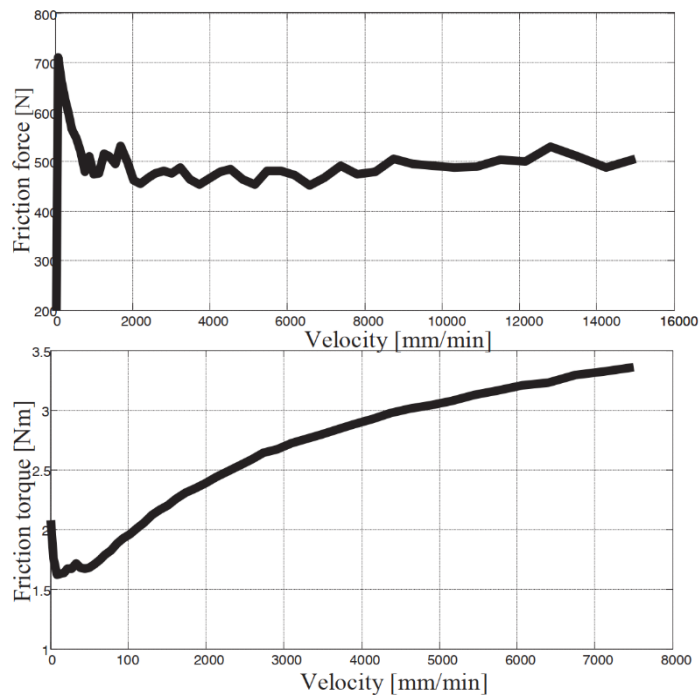

Figure 44 Typical friction behavior of feed axis (a) linear motor driven

(top); (b) ball screw driven (down) [125]

Liu et al. [141] established a state-dependent friction model considering time-varying factors (feed speed and dual-drive structure) through parameter identification experiments. They found that feed system friction is influenced by time-varying factors like feed speed and experimentally verified the model's effectiveness. They discovered that the feed system's natural frequency changes under different motion states.

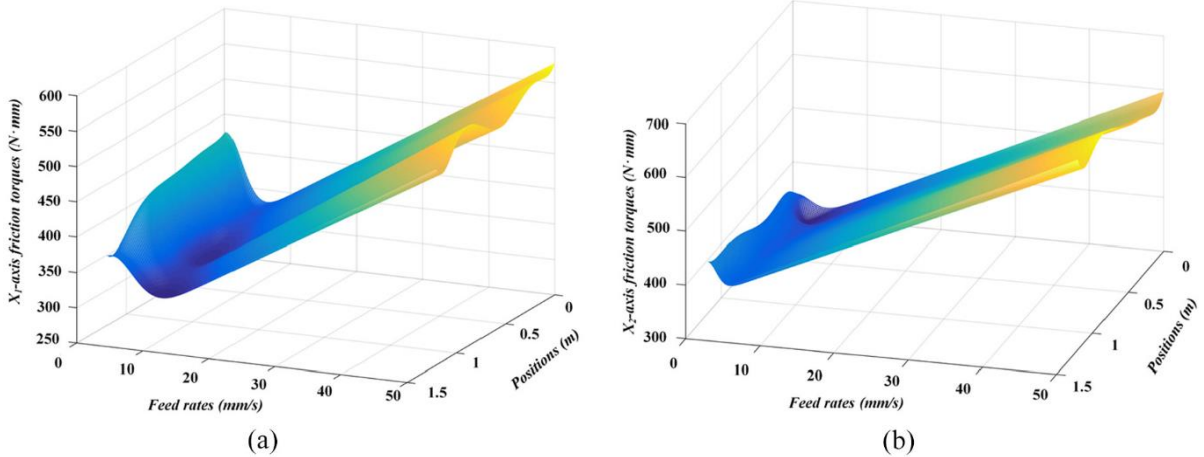

Figure 45 Friction torques with feed rates: (a) X1-axis and (b) X2-axis. [141]

Zhang et al. [142] proposed a method for variable parameter modeling of the ball screw feed system dynamics considering the influence of the moving process. The effectiveness o was validated through experiments. During the research, it was found that the natural frequency of the feed system changes under different motion conditions Zhang et al. [143] established a dynamics model of the ball screw feed system considering the effect of feedrate. The results have shown that the natural frequency of the system in motion is larger than that in a stationary state. Further, it also varied depending on the federate, as shown in Figure 47 and Figure 48.

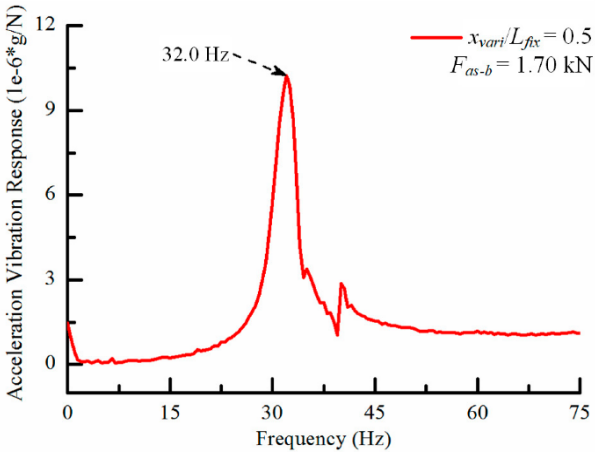

(a) static state

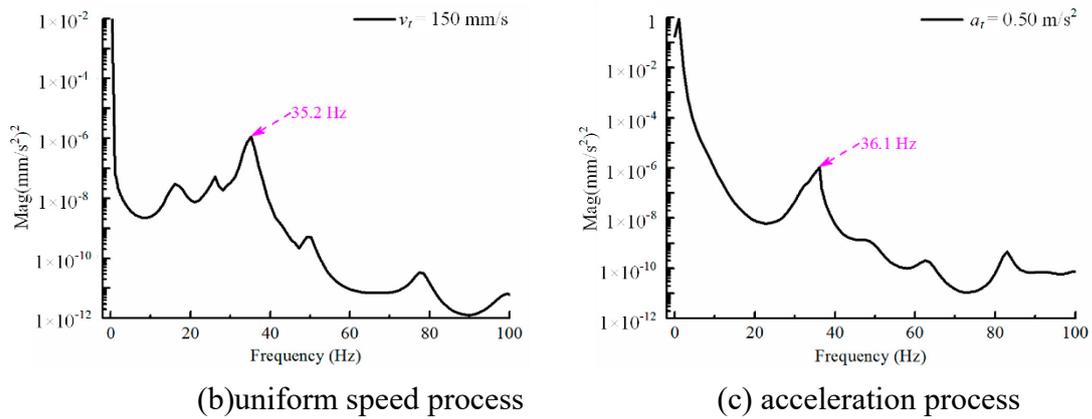

(b) uniform speed process  (c) acceleration process

Figure 46 Frequency response of the feed system in different states [142]

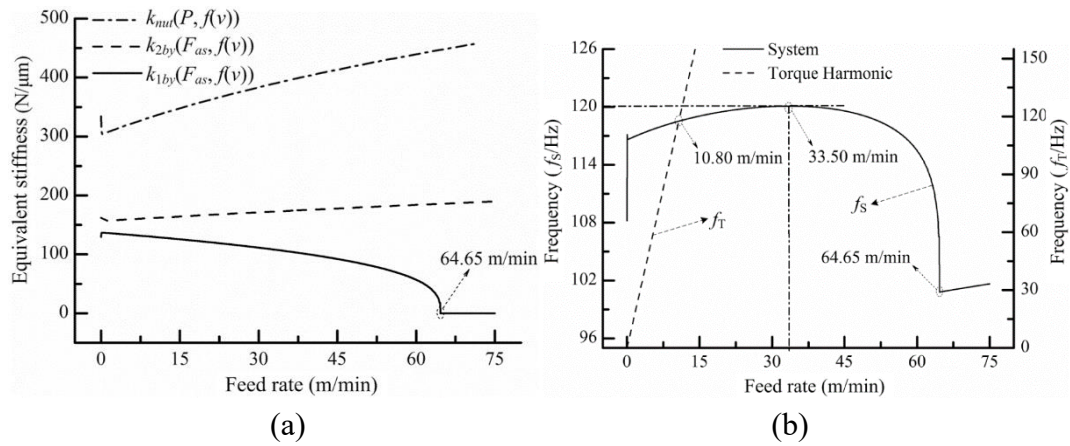

(a)  (b)

Figure 47 The variation in the equivalent stiffness of kinematic joints(a) and frequency(b) with feed rates[143]

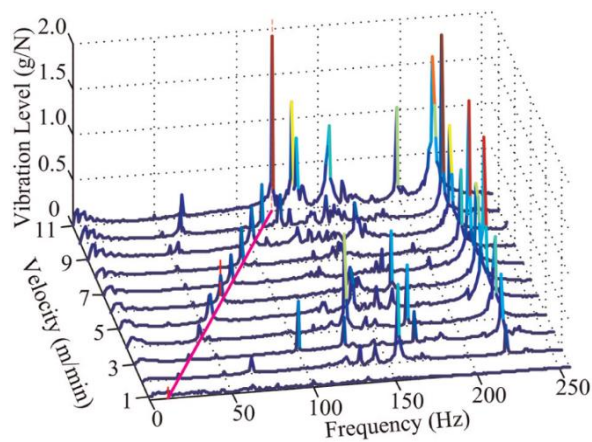

Figure 48 The tested system frequency response at different feed rates[143]

Lu et al. [144] developed a dual-drive feed mechanism dynamics model considering the effects of feedrate and moving part positions. The results show that the natural frequency is greater in the motion than in the static state. Moreover, its behavior varied depending on the feedrate and moving part positions.

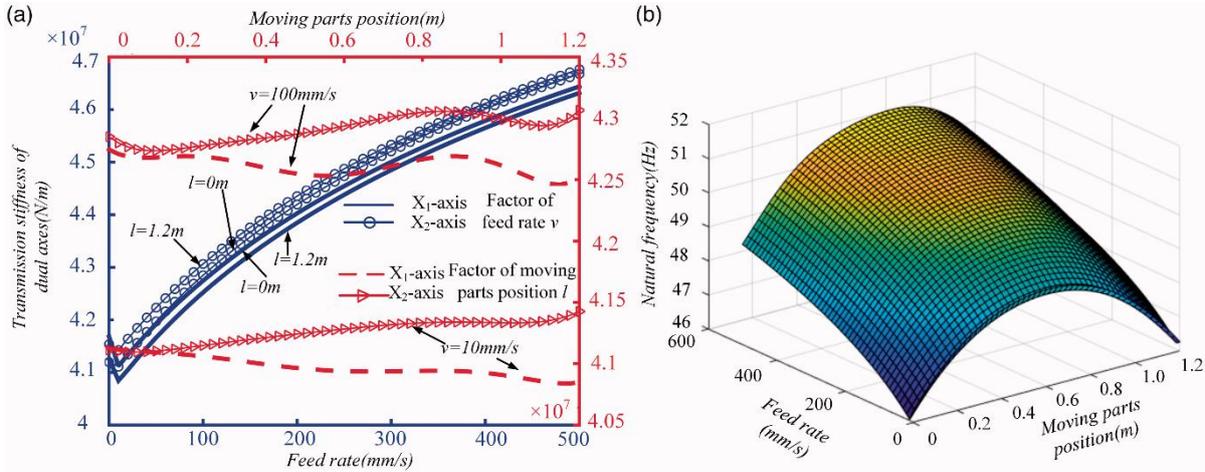

Figure 49 Variation of transmission stiffness and natural frequency. (a) Variation of $X_1$-axis and $X_2$-axis transmission stiffness with feed rates and moving parts positions; (b) The first natural frequency of the dual-drive feed mechanism. [144]

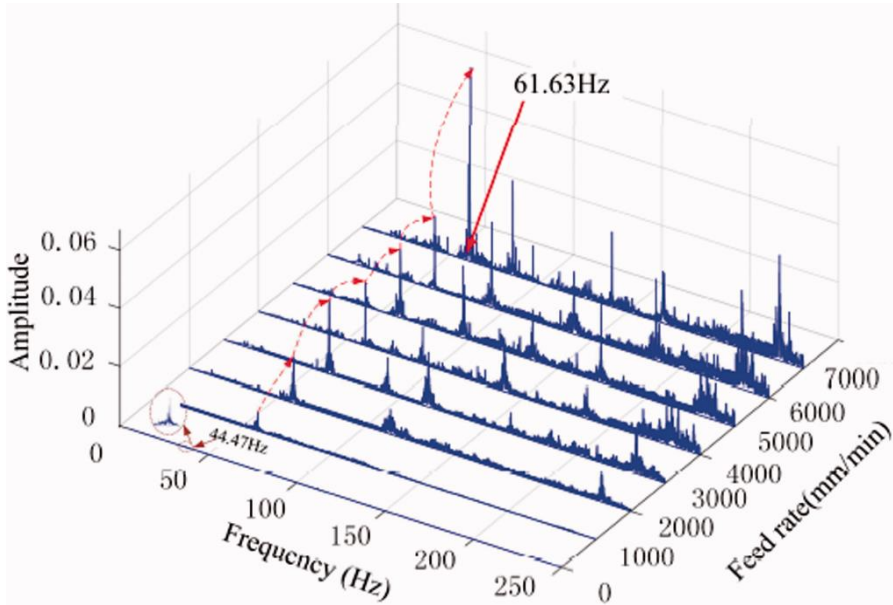

Figure 50 Vibration characteristics of the dual-drive feed mechanism with feed rates[144]

Additionally, different motion processes lead to variations in inertial forces. The variation in inertial forces affect the system through vibrations and deformations during the machining process at high speeds using high-precision machine tools. Thus, the dynamic performance of the machine tool is also affected. [145]

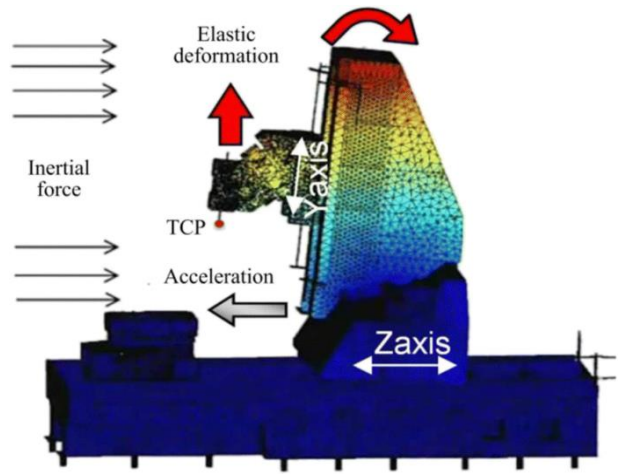

Figure 51 Deformation of mechanical series outside the servo loop caused by inertial force[44]

All the referenced studies show that the change in feedrate affects the dynamic performance of the system – its change will also change the inertia, bond surface characteristics, and friction of the system. This will, in turn, affect the mechanical characteristics of the system, such as vibration condition, natural frequency, and deformation. Finally, it is also necessary to study how the coupling effect will change after considering the influence of the motor and control system.

# 4 Research on dynamically matched design methods

Comprehensive consideration of subsystem interactions in the dynamic optimization design of CNC machine tools is essential to achieve optimal matching among subsystems and maximize dynamic performance. Existing design methods for feed system dynamics, such as inertia matching and optimization methods, have limitations. Inertia matching does not clarify subsystem coupling mechanisms and is limited to simple matching of motor and load inertia. These approaches often fail to directly indicate whether the design can meet the dimensional accuracy and surface quality requirement. Optimization methods, although considering the integrated effects of subsystems, are computationally complex and lack clear understanding of underlying subsystem coupling mechanisms, limiting practical engineering application.

## 4.1 Inertia matching method

The inertia matching method between motor and load is commonly used in the dynamics matching design of CNC machine tool feed systems[26]. Dequidt et al. [146] pointed out that the coupling effect between mechanical vibration and controller must be considered when determining the inertia ratio. Caracciolo et al. [25] limited the inertia ratio to a pre-defined range and included the constraints of mechanical, motor, and control systems for ball screw integration design. That way, the servo system can achieve the specified dynamic performance.

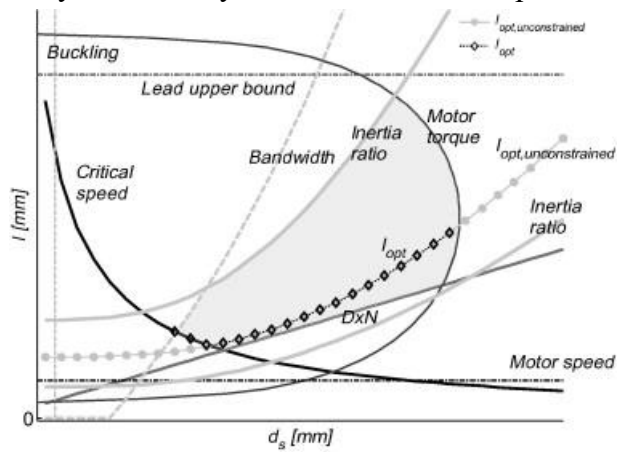

Figure 52 Allowable domain and optimal lead [25]

Next, Kong et al. [147] studied the inertia matching problem of a five-degree-of-freedom manipulator, mainly analyzing the relationship between the end output energy and inertia ratio. The relationship between the two was analyzed, and it was concluded that the end output energy of the manipulator with an inertia ratio between 0.4 and 2.6 was higher. However, the above-referenced study did not consider how the change of motion process affects the coupling effect of the electromechanical system, as well as the choice of the inertia ratio. The inertia matching of the motor and mechanical system to design high-speed and high-precision CNC machine tool feed systems is more complex [148].

Zhang et al. [37] used an intelligent optimization algorithm to exclude the influence of control parameters on the dynamic performance of the servo control system. The authors then studied the influence of the moving process and inertia ratio on the dynamic performance of the system. In the high-speed and high-precision CNC machine tool feed systems design, the inertia ratio should be strictly achieved within a certain range, neither too large nor too small.

The design method based on the inertia ratio is straightforward and easy to implement in engineering practice. However, it cannot be quantified to determine the inertia ratio that can make the system achieve optimal performance. Furthermore, the underlying mechanism of inertia ratio matching design method should also be studied.

## 4.2 Optimization algorithm

This integrated design scheme of the feed system based on a multi-objective optimization method takes the system dynamic performance as the objective function. It optimizes the design of parameters of several sub-systems. Such methods establish a complete process for the integrated optimal design of CNC machine tool feed systems.Such processes generally optimize each parameter through mathematical methods to ensure optimal performance. The implementation process is slightly more complex and demanding for use in engineering practice. Most engineers only consider matching the motor and mechanical physical entities to provide easy and fast design.

The nonlinear multi-objective optimization method was used to integrate the mechatronic system design in several studies [149–154]. The general process is as follows: first, to establish a complete dynamic system model. Then the constraints and objective function are determined according to the design objectives. The objective function generally contains indicators characterizing the dynamic accuracy and system response speed. Finally the design parameters are optimized to achieve the optimal dynamic performance of the feed system.

Kim et al. [129,155–157] proposed a feed system design method that considers the coupling effect of the mechanical system and control system and established a nonlinear multi-objective optimization design method. The objective function reflected the dynamic accuracy and response speed of the feed system, ensuring the dynamic performance and feed system stability (among other constraints) while the design parameters were optimized.

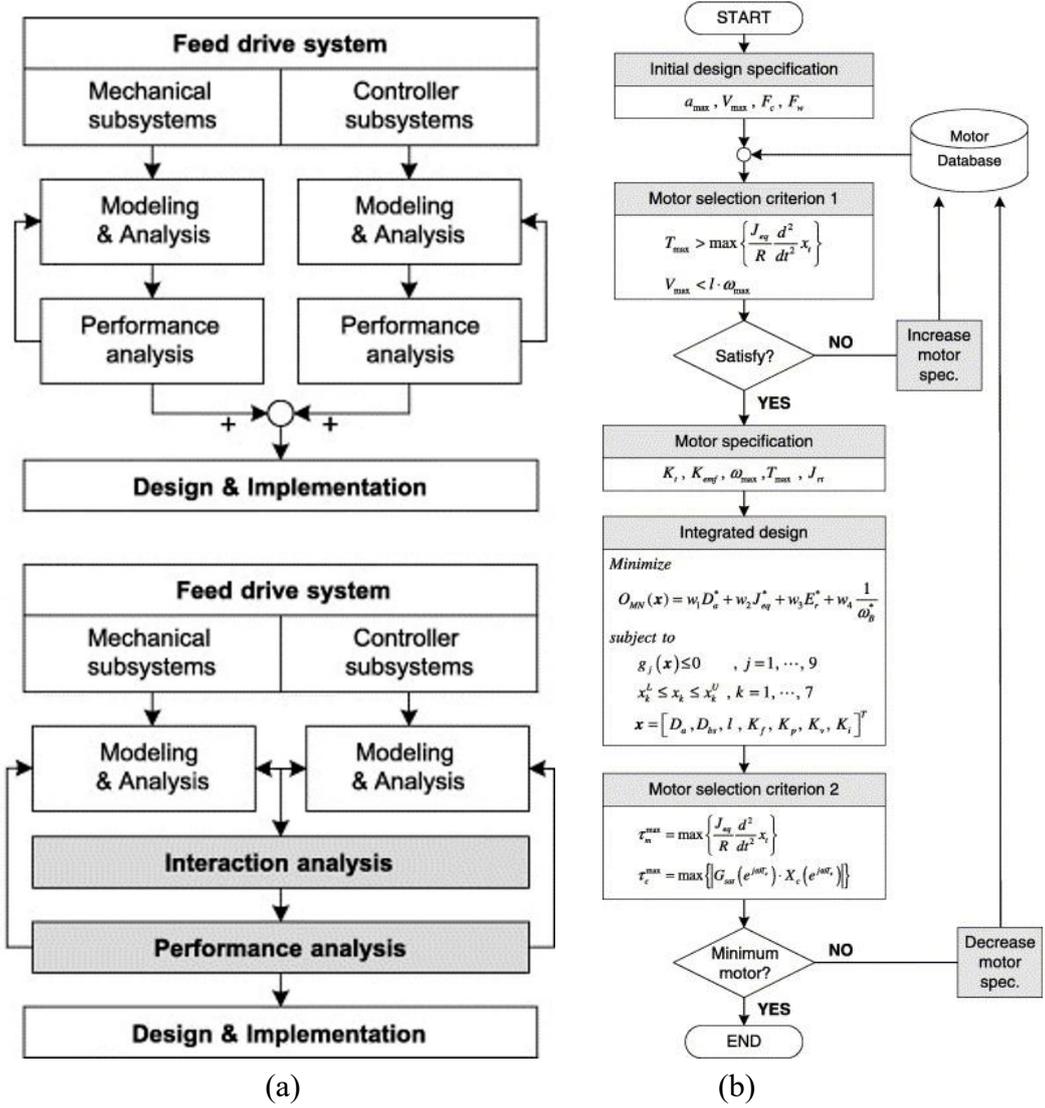

Figure 53 (a) Component design vs. systematic design approaches (b) Flow chart of Kim's integrated design procedure [129,155–157]

Further, some researchers also studied the problem from an analytical point of view [158–162], considering the torque, peak torque, and electric energy to match the motor and the load drive system optimally. Yong et al. [163] optimally adjusted the control parameters of the control sub-system according to mechanical characteristics, decreasing the rise time and stability time of its step response from 0.089 s and 0.104 s (before the optimization), respectively, to 0.038 s and 0.092 s. The dynamic response performance was significantly improved.

# 5 Summary and Outlook

Over the past decades, significant efforts have been made in the analysis and design of CNC machine tool feed system dynamics, resulting in clearer understanding of influencing mechanisms and more advanced design schemes. This paper reviews research on optimization design methods and dynamic performance mechanisms of CNC machine tools.

In summary, this paper explores potential future development directions in CNC machine tool design:

(1) Define the required dynamic characteristics of the feed system and establish dynamic performance indicators. Traditional design approaches struggle to determine if accuracy requirements are met during the design phase. Feed system design should clarify required dynamic characteristics. Based on machining scenario needs, with the first-order mode frequency of the mechanical structure meeting requirements, design should be conducted using position error $\Delta e(t)$ and their derivatives as dynamic performance indicators to ensure the feed system meets required dynamic performance, ensuring dimensional accuracy and surface quality of machined workpieces.

(2) Optimal subsystem performance does not guarantee optimal overall feed system performance. The performance of CNC machine tool feed systems is influenced by the integrated effects of subsystem coupling. Comprehensive research on these influence mechanisms is essential to guide optimization design. Current research on subsystem coupling mechanisms focuses on specific interactions, such as control parameter tuning considering mechanical characteristics, inertia matching between motor and mechanical load, and the impact of motion processes on mechanical system interface. Comprehensive research encompassing all subsystem interactions is lacking.

(3) Dynamic optimization design of CNC machine tool feed systems should consider complex subsystem interactions and their integrated impact on dynamic performance. Design should aim to meet dynamic performance indicators outlined in point (1), enabling precise quantitative design and ensuring the feed system meets dynamic performance requirements of machining scenarios, guaranteeing dimensional accuracy and surface quality of machined workpieces.